\documentclass[preprint,11pt,authoryear]{elsarticle}

\usepackage{amssymb}
\usepackage{amsmath}
\usepackage{booktabs}
\usepackage{tikz}
\usepackage[linesnumbered,ruled,vlined]{algorithm2e}
\usepackage{caption}
\usepackage{subcaption}
\usepackage{float}
\usepackage{comment}
\usepackage{fullpage}
\usepackage[percent]{overpic}

\journal{arXiv}

\begin{document}

\begin{frontmatter}

\title{oRANS: Online optimisation of RANS machine learning models with embedded high-fidelity data generation}

\author[ox]{D. Dehtyriov\corref{cor1}} 
\ead{daniel.dehtyriov@maths.ox.ac.uk}
\cortext[cor1]{Corresponding author}
\author[nd]{J. F. MacArt}
\ead{jmacart@nd.edu}
\author[ox]{J. Sirignano}
\ead{justin.sirignano@maths.ox.ac.uk}

\affiliation[ox]{organization={Mathematical Institute, University of Oxford},
            addressline={Andrew Wiles Building, Woodstock Rd}, 
            city={Oxford},
            postcode={OX2 6GG}, 
            country={United Kingdom}}

\affiliation[nd]{organization={Department of Aerospace and Mechanical Engineering, University of Notre Dame},
            addressline={369 Fitzpatrick Hall of Engineering}, 
            city={Notre Dame},
            postcode={IN 46556},
            country={USA}}

\begin{abstract}
Deep learning (DL) has demonstrated promise for accelerating and enhancing the accuracy of flow physics simulations, but progress is constrained by the scarcity of high-fidelity training data, which is costly to generate and inherently limited to a small set of flow conditions. Consequently, closures trained in the conventional offline paradigm tend to overfit and fail to generalise to new regimes.
We introduce an online optimisation framework for DL-based Reynolds-averaged Navier--Stokes (RANS) closures which seeks to address the challenge of limited high-fidelity datasets. Training data is dynamically generated by a simulation at direct numerical simulation (DNS) resolution embedded within a subdomain of the RANS domain. The RANS solution supplies boundary conditions to this embedded simulation, which in turn provides mean velocity and turbulence statistics that are used to update a DL closure model during the simulation. This feedback loop enables the closure to adapt to the target flow of the embedded simulation, avoiding reliance on precomputed datasets and improving out-of-distribution performance.
The approach is developed on the stochastically forced Burgers equation and on turbulent channel flow at $Re_\tau=180$, 270, 395 and 590 with embedded domain lengths $1\leq L_0/L\leq 8$, and is then applied to spanwise-rotating channel flow at rotation numbers $Ro_b=0.15$ and $0.5$, a flow in which the Coriolis force vanishes from the equations of a linear eddy-viscosity closure, which is therefore blind to rotation. The learned coefficients are interpretable, recovering the near-wall damping of low-Reynolds-number models in the channel and, under rotation, an eddy viscosity that falls to the order of the kinematic viscosity on the stabilised side and the production that rotation sustains in the core. Online-optimised RANS models generally outperform both offline-trained and literature-calibrated closures, with accurate training achieved using modest embedded subdomains. In the rotating channel they recover the rotation-induced asymmetry from non-rotating inflow turbulence and a rotation-blind closure. Performance degrades primarily when boundary-condition contamination dominates or when domains are too short to capture low-wavenumber modes. This framework provides a route to physics-informed machine learning closures, enabling data-adaptive reduced-order models that adapt to each flow regime without requiring large precomputed training datasets.
\end{abstract}


\end{frontmatter}

\section{Introduction}\label{sec:introduction}

Fluid turbulence in the continuum flow regime is fully described by the Navier--Stokes equations. Solving these equations exactly, i.e., direct numerical simulation (DNS), in flow regimes of engineering interest is typically infeasible due to the large range of spatiotemporal scales required to accurately resolve the nonlinear physics. To remain computationally feasible, simulations often reduce the necessary spatiotemporal resolution via large eddy simulation (LES) using spatial filtering or Reynolds-averaged Navier--Stokes (RANS) simulations using spatiotemporal averaging.
Both require physical approximations  to be introduced via turbulence closure models.
\par
High-fidelity simulation data has historically played a pivotal role in calibrating turbulence closure models. Many classical LES/RANS model parameters (e.g., the Kolmogorov constants, Smagorinsky coefficient or damping functions for near-wall behaviour) have been informed by matching DNS or experimental benchmarks \citep{Launder1974, Wilcox1998, Smagorinsky1963}. The availability of high-fidelity data has thus opened the door to data-driven turbulence modelling, wherein one uses measurements or simulations of the flow field to guide the form or parameters of the closure. A fundamental challenge, especially for deep learning closure models with large numbers of parameters, is the limited number of high-fidelity datasets which are typically available for calibrating closure models. 

\subsection{Data driven turbulence modelling}

The Reynolds-averaged Navier--Stokes equations, which solve only for the mean flow variables, remain in widespread use due to their low computational cost \citep{pope2000}. This comes at the cost of an unclosed term with significant complexity, representing the effect of the fluctuating field on the mean flow. The time averaging shifts the challenge from the large computational effort required for solving the instantaneous equations to modelling the flow physics embedded in the unclosed RANS equations.
\par
Conventional RANS models solve additional transport equations for the unclosed term, which themselves depend on closure coefficients. These coefficients are typically calibrated using data from canonical flows and based on asymptotic arguments \citep{Hanjalic1972,Menter1994,Wallin2000}, allowing for tractable `plug-and-play' solutions for arbitrary engineering flows of interest. Despite the widespread use of RANS however, it is well known that its predictive accuracy is poor in flows with strong anisotropy, separation or curvature, where the underlying assumptions break down \citep{Spalart2000}. As RANS closures are tuned to a narrow set of canonical flows, they lack universal accuracy across a broad spectrum of turbulent flow configurations. This lack of universality has motivated extensive research into improved closure strategies, both through physics-based reasoning and, more recently, data-driven approaches.
\par
A seminal contribution in this direction was the Tensor Basis Neural Network (TBNN) proposed by \citet{Ling2016}, which imposed Galilean invariance through a custom multiplicative layer to learn nonlinear mappings from local flow features to the anisotropy tensor. \citet{Wang2017} introduced an alternative strategy that learned the discrepancy between RANS-predicted and DNS-derived Reynolds stresses, enabling data-informed corrections to classical models. \citet{Parish2016} developed the Field Inversion and Machine Learning (FIML) framework, in which a spatially distributed modification to a RANS closure is inferred via inverse modelling and then generalised through supervised learning. Such approaches demonstrated that augmenting eddy-viscosity models with data-driven corrections can significantly improve RANS predictions for flows similar to the calibration cases. Neural-network subgrid-scale models have been trained in the same way for two-dimensional turbulence \citep{Maulik2019} and for channel flow \citep{Park2021}, in each case on filtered DNS of the flow to be modelled.
\par
Both DNS and well-resolved large-eddy simulation (LES) have provided detailed turbulence statistics that were historically unattainable from experiments alone. For instance, the DNS of fully developed channel flow \citep{Kim1987} resolved all essential scales of near-wall turbulence and reported a comprehensive set of statistics for comparison with experiments. These high-fidelity datasets may inform the physics of traditional RANS models, and they serve as a reference for developing and calibrating new models. 
\par
However, limited data diversity leads to overfitting of the learned model to the calibration flows, yielding poor generalisation to new regimes \citep{Duraisamy2021}. Offline-trained ML models often encode strong priors based on the training set and exhibit degraded performance when applied to flows with different geometries, Reynolds numbers, or dominant physics. This generalisation gap has spurred efforts to regularise models using physics-informed features, invariant bases, or sparsity-promoting architectures. Recent reviews \citep{Duraisamy2019, Brunton2020} have emphasised the importance of embedding physical constraints into ML turbulence models to ensure robustness and extrapolative power. One response is to assemble the training set from canonical building-block flows chosen to span the regimes of interest \citep{LozanoDuran2023}, which extends the coverage of the data but still requires the relevant flow to be represented in it. Probabilistic learning approaches have also been introduced to provide uncertainty quantification (UQ) in ML-predicted closures. For instance, the Reynolds stress prediction can be formulated as a probabilistic mapping, allowing confidence intervals to be estimated alongside mean predictions \citep{Xiao2019}.
\par
More recent data-driven approaches have sought to directly embed machine-learned closures into the RANS or LES equations and optimise them against high-fidelity data. In these PDE-constrained formulations, the functional form of the unresolved terms is represented by a flexible model (such as a neural network), and its parameters are adjusted by requiring the RANS/LES solution to match reference data. Adjoint methods are typically employed to compute the gradient of the loss with respect to the closure parameters. \citet{Sirignano2023_pde} provide a rigorous convergence analysis of this approach for a model elliptic PDE system, with adjoint-based optimisation then used to train a neural network functioning as a RANS closure model, calibrating it on several DNS datasets of turbulent channel flow. Similarly, \citet{Sirignano_2023_DLLES} developed a deep-learning LES subgrid closure by directly matching filtered DNS data for flow around various bluff bodies, and \citet{List2022} train a closure through a differentiable solver over many time steps, which likewise optimises the model through the governing equations on precomputed data. \cite{Bae-2022}, \cite{Zhou2022}, and \cite{Vadrot-2023} developed reinforcement learning methods to train wall models for LES. These examples underscore the potential of offline-trained deep-learning closures: when provided with sufficient high-fidelity data of a given flow class, the ML-based models can encode complex turbulent transport physics and improve upon conventional closures.
\par
In parallel to these ML-driven strategies, non-ML approaches have also been developed to reduce the cost of incorporating high-fidelity information into RANS and LES closures. One class of methods is embedded DNS frameworks \citep{He2018, Chen2022, Chen2023, He2023}, where local fine-mesh DNS blocks are coupled to a global coarse-mesh domain through block-spectral mappings and source terms, reducing mesh-count scaling with Reynolds number compared to conventional LES or DNS. In these methods the fine-mesh block resolves the near-wall turbulence while the coarse-mesh solution supplies the large-scale structures that act on it, and neither a spectral gap nor a separation of scales is assumed \citep{Chen2022}. Nested LES, in which a fine-mesh inner domain is coupled to a coarse outer one, has been applied to channel flow in the same spirit \citep{Tang2016}. The question these methods raise, which scales a local fine-mesh region must resolve for the coarse solution outside it to be correct, is the question the present method poses for a learned closure. A complementary line of work focuses on boundary condition generation, where synthetic inflow turbulence methods \citep{Klein2003, Hao2022, Dreze2023} generate realistic inflow statistics and correlations, reducing the domain length required to achieve fully developed turbulence. Both approaches illustrate how embedding or inflow strategies can lower the computational burden of integrating high-fidelity information into turbulence simulations.
\par
Despite these advances, a central limitation of ML-based closures remains their reliance on offline training with precomputed high-fidelity datasets. We use the term offline for any training on precomputed data, including the PDE-constrained training described above \citep{Sirignano2023_pde,Sirignano_2023_DLLES,List2022}, which is sometimes referred to as \emph{a posteriori} training because the closure is optimised through the governing equations. By online training we mean that the training data are themselves generated during the simulation. Offline models are constrained by dataset availability, limited flow diversity, and the attendant risk of overfitting. Preliminary work has explored online optimisation of LES closures using embedded DNS  \citep{Sirignano2023_OnlineDL}.

\subsection{Motivation}

Areas such as computer vision and natural language processing have seen rapid progress in machine learning, in large part due to the availability of vast, high-quality datasets. In contrast, scientific applications, such as turbulence modelling, lack such data abundance. Turbulence data is typically generated via experiments or high-fidelity direct numerical simulation, both of which are expensive and limited to a finite set of fixed geometries or Reynolds numbers. DNS is computationally prohibitive at high Reynolds numbers, while experimental campaigns are constrained by cost, facility availability, and difficulties in measuring three-dimensional, time-resolved fields. As a result, real-world engineering applications suffer from data sparsity.
\par
In a typical \textit{offline supervised learning} workflow, this limited data is used to train the parameters $\theta$ of a turbulence model, which is then deployed to new regimes without further adaptation. This often results in reduced accuracy, as the model must extrapolate beyond its training distribution. This traditional workflow proceeds as follows:
\begin{enumerate}
\item Generate DNS data for a finite set of conditions
\begin{align}
\frac{\partial v^{\textrm{DNS}}}{\partial t} &= \mathcal{F}(v^{\textrm{DNS}}; \ \lambda), \ x \in \Omega,
\end{align}
where \( \mathcal{F} \) represents the Navier--Stokes operator (or other nonlinear operator),  $\lambda$ collects the conditions (Reynolds number, geometry, boundary conditions, forcing, etc.), and $\Omega\in\mathbb{R}^{d}$ is the $d$-dimensional Cartesian domain. Additional constraints such as the incompressibility condition
$0 = \nabla \cdot v^{\textrm{DNS}}$
can also be imposed.

\item Train a closure model $h_{ij}$ incorporated in the RANS/LES partial differential equation (PDE),
\begin{align}
\frac{\partial \overline{v_{\theta}}}{\partial t} &= \mathcal{F}\left(\overline{v_{\theta}} ;\ \lambda \right) - \frac{\partial h_{ij}}{\partial x_j}(\nabla \overline{v_{\theta}}; \ \theta), \ x \in \Omega,
\end{align}
where the overbar $\overline{\cdot}$ represents a spatiotemporal average/filtering operation, and $h_{ij}$ is the learned closure parametrised by $\theta$. Additional constraints can likewise be imposed,
\begin{align}
0 &= \nabla \cdot \overline{v_{\theta}}.
\end{align}
Because $\overline{v}$ variables are filtered/averaged quantities, $v^\textrm{DNS}$ must be filtered consistently before comparison. The model is trained by minimizing the discrepancy with the dataset:
\begin{equation}
    L(\theta) = \int_0^T\int_\Omega|| \overline{v_{\theta}} - \overline{v^\textrm{DNS}} ||^2  \ dx \ dt,
\end{equation}
where $T$ denotes the training time window.
\item Make out-of-sample predictions for unseen  conditions not included in $\lambda$.

\end{enumerate}

\par
Despite recent advances in deep learning for turbulence modelling, this \textit{offline paradigm} remains fundamentally limited by the scope of the available high-fidelity training data, which is costly to generate and inherently restricted to a finite set of flow regimes. As a result, the trained model lacks the capacity to adapt when applied to flow conditions that lie outside the support of the training distribution, often leading to degraded predictions and a failure to generalise in physically meaningful ways. In the spanwise-rotating channel of section~\ref{sec:rotating}, for example, a standard linear eddy-viscosity closure contains no term through which system rotation can act, so no offline calibration on non-rotating data can produce the rotation response.
\par
To address this limitation, we propose to train turbulence closures using an \textit{embedded, online learning} approach, in which the closure is optimised through the governing equations, as in PDE-constrained offline training, and the data on which it is trained are in addition generated during the flow simulation itself. A high-fidelity simulation at DNS resolution, referred to in the literature as an embedded DNS, is embedded within a subdomain of a larger RANS simulation. The RANS solution provides boundary conditions to the embedded simulation, which in turn supplies the high-fidelity flow statistics needed to update the turbulence model parameters.

\begin{figure}
    \centering
    \includegraphics{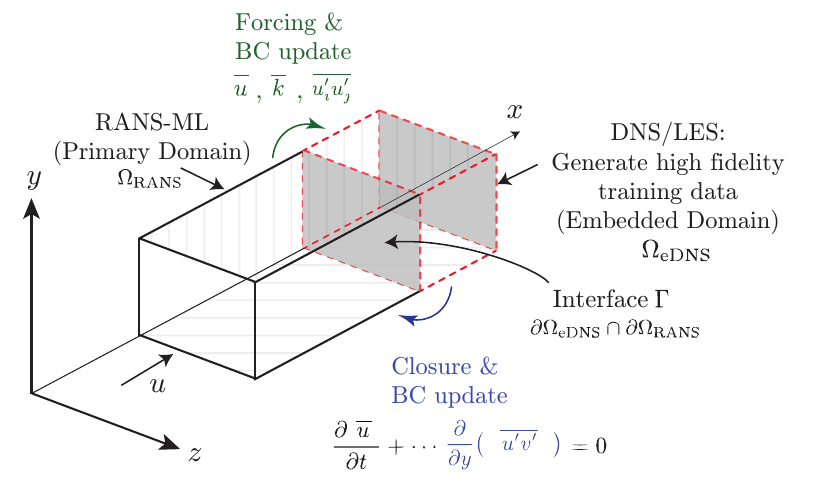}
    \caption{Schematic of the coupled RANS and embedded simulation framework. The primary RANS-ML domain provides boundary conditions and forcing (mean velocity, turbulent kinetic energy, Reynolds stresses) to the embedded high-resolution region, which in turn supplies high-fidelity data to update the RANS closure and boundary conditions. This coupling enables consistent training of turbulence models on statistically representative flow fields.}
    \label{fig:schematic}
\end{figure}

This feedback loop enables the model to adapt during the simulation, reducing the reliance on offline data and improving predictive accuracy across both domains. A schematic of the setup is shown in Figure~\ref{fig:schematic}. The embedded, online learning workflow proceeds as follows:
\begin{enumerate}

\item Simulate coupled high- and low-fidelity PDEs on two domains, where $\Omega_{\mathrm{eDNS}}$ denotes the embedded, high-fidelity (eDNS) subdomain, and $\Omega_{\mathrm{RANS}}$ denotes the surrounding low-fidelity RANS domain, with interface
$\Gamma=\partial\Omega_{\mathrm{eDNS}}\cap\partial\Omega_{\mathrm{RANS}}$:
\begin{align}
\frac{\partial u^{\mathrm{eDNS}}}{\partial t} &= \mathcal{F}\!\left(u^{\mathrm{eDNS}}\right), 
\ \  x \in \Omega_{\mathrm{eDNS}}, \label{eq:coupledpde1}\\
0 &= \nabla \cdot u^{\mathrm{eDNS}}, \\
0 &= 
\mathcal{F}\!\left(\overline{u}_{\theta(t)}^{\mathrm{RANS}}\right) 
+ \nabla \cdot h\!\left(\nabla \overline{u}_{\theta(t)}^{\mathrm{RANS}};\,\theta(t)\right),
\ \  x \in \Omega_{\mathrm{RANS}}, \\
0 &= \nabla \cdot \overline{u}^{\mathrm{RANS}}_{\theta(t)}, \\
u^\mathrm{eDNS} \big|_{\Gamma} &= \mathcal{T}\!\left(\overline{u}_{\theta(t)}^{\mathrm{RANS}}\big|_{\Gamma};\,\theta(t)\right), 
\ \  x \in \Gamma. \label{eq:coupledpde2}
\end{align}
Here $u=(u_1, \ldots , u_d)$ denotes the $d$-dimensional velocity vector (where $d = 1, 2,$ or $3$), and 
$h(\cdot \ ;\theta)$ is a learned closure model producing a vector field whose 
divergence modifies the RANS momentum equation. We define the composite field $u$ by
$u=u^{\mathrm{eDNS}}$ on $\Omega_{\mathrm{eDNS}}$ and
$u=\overline{u}^{\mathrm{RANS}}$ on $\Omega_{\mathrm{RANS}}$, where $\Omega_{\mathrm{eDNS}}\in\mathbb{R}^d$, $\Omega_{\mathrm{RANS}}\in\mathbb{R}^p$, and $p\leq d$ depending on the number of statistically homogeneous dimensions.
The operator $\mathcal{T}$ supplies boundary data to the embedded subdomain from the surrounding RANS solution at the interface $\Gamma$.
In the simplest case, $\mathcal{T}$ is the identity, directly imposing the RANS field on $\Gamma$. In this work, $\mathcal{T}$ is time dependent and augments the RANS state with rescaled fluctuations to provide statistically consistent turbulent inflow (see section~\ref{sec:oranssetup} for details). Note that $u^{\mathrm{eDNS}}$ depends on $\theta$ only through $\mathcal{T}$ on $\Gamma$; the interior operator $\mathcal{F}$ does not depend explicitly on $\theta$.

\item Continuously update the model parameters and boundary conditions for asymptotic minimisation of the closure-modelling error:
\begin{equation}\label{eq:gradientflow}
\frac{d\theta}{dt}
= \alpha\int_{\Omega_{\mathrm{eDNS}}}
\big(u^{\mathrm{eDNS}} - \overline{v}_{\theta(t)}^{\mathrm{RANS}}\big)\,
\nabla_{\theta}\,\overline{v}_{\theta(t)}^{\mathrm{RANS}}\, dx,
\qquad \theta(0)=\theta_0,
\end{equation}
where $\alpha$ is the learning rate, and where $\bar v_\theta^{\mathrm{RANS}}$ denotes the RANS surrogate field solved on the entire domain $\Omega$. This is necessary because $\overline{u}^{\mathrm{RANS}}$ is only defined on $\Omega_{\mathrm{RANS}}$, while the parameter update in \eqref{eq:gradientflow} is evaluated over $\Omega_{\mathrm{eDNS}}$. When the solution has statistically homogeneous dimensions, as in the channel flow test case considered herein, one may equivalently use $\overline{u}^{\mathrm{RANS}}$ in place of $\overline{v}^{\mathrm{RANS}}$ for computational efficiency. We adopt this convention throughout the remainder of the paper.
\end{enumerate}
By comparison to the {offline supervised learning} approach, our framework generates training data directly from the exact physical conditions and geometries on which predictions are desired. This enables model training in computationally challenging flow regimes without the limitations of dataset sparsity and overfitting, since the reduced-order model is trained locally on the embedded-simulation data but is then applied to the surrounding RANS domain to represent the unresolved dynamics, thereby generalizing to the remainder of the flow field outside the high-fidelity region. The present formulation can be readily extended to arbitrary closures such as the $k$--$\epsilon$ and $k$--$\omega$ models. More broadly, the strategy applies to any nonlinear PDE system where simplification (e.g., temporal averaging or spatial filtering) introduces unclosed terms.

\subsection{Paper outline}

We develop an online optimisation method for RANS ML closure models to address challenges with overfitting to limited datasets, where the ML closure model is continuously updated during the simulation based on data from the evolving embedded high-fidelity flow field. This approach enables the closure model to adapt to the specific configuration being simulated, potentially overcoming the generalisation problem inherent to offline-trained ML closure models.

\par
A fully online-trained RANS closure framework is developed that uses an embedded high-resolution subdomain to iteratively correct the closure model during the simulation itself. Importantly, this implies that the closure is trained directly on the geometry and physics of the target simulation, eliminating the mismatch between training and deployment. In this framework, high-fidelity regions within the RANS domain are simulated at DNS resolution, and their time-averaged quantities are used to compute a local loss. This loss is minimised via stochastic gradient descent to update the parameters of a neural network closure model embedded in the RANS solver. The result is a data-adaptive closure that evolves with the flow and corrects itself \textit{in situ}.
\par
The present study develops this methodology in a canonical setting, but the framework is general and extensible to other closures and nonlinear PDEs with unclosed terms. Existing deep-learning closures, including those optimised through the governing equations \citep{List2022,Sirignano_2023_DLLES}, are trained on data computed beforehand and are limited to the flows for which such data exist. Here the training data are generated during the simulation, on the target flow, by an embedded region whose boundary conditions are supplied by the closure being trained, so that no precomputed data of the target flow are required and the closure can acquire physics that neither its equations nor the available data contain. The method extends the preliminary online training of LES closures with embedded DNS \citep{Sirignano2023_OnlineDL} to RANS closures trained on time-averaged statistics. In contrast to embedded-DNS methods \citep{He2018,Chen2022}, which estimate the closure from the fine-mesh blocks without a learned model and retain the blocks in every computation, the high-fidelity region here trains the model that supplies its boundary conditions, and once trained the model can be used without it. 
In this arrangement the RANS closure carries the large-scale statistics of the flow, and the embedded region resolves only the small and moderate scales, driven at its inlet by the mean velocity and the energy of the RANS solution. Sections~\ref{sec:nsresults} and~\ref{sec:rotating} examine what the region must regenerate for the closure trained on it to be accurate, how short the region can be, and whether it can supply a response, that to rotation, which the closure itself does not contain.
Our contributions include:
\begin{enumerate}
    \item The formulation of \textbf{online-optimised RANS (oRANS)}, a coupled RANS/embedded high-resolution simulation framework with continuous, online training of ML closures on the data generated in the embedded region;
    \item The derivation of the \textbf{discrete adjoint of the ML-augmented $k$--$\omega$ turbulence model} for PDE-constrained optimisation, together with an efficient stencil-wise implementation;
    \item The development of an \textbf{inflow rescaling procedure} that enables statistically representative embedded simulations without requiring long periodic boxes;
    \item The application to \textbf{spanwise-rotating channel flow}, in which the closure recovers a rotation response that is absent from both the baseline closure and the non-rotating data available offline.
\end{enumerate}

The paper is organised as follows. Section 2 introduces the oRANS algorithm in the setting of conservative PDEs, which provides the working formulation for the three systems studied here: stochastically forced Burgers’ equation, incompressible turbulent channel flow and spanwise-rotating channel flow, and presents an efficient reverse-mode adjoint formulation. Section 3 validates the framework on the stochastic Burgers equation. Section 4 applies oRANS to the Navier--Stokes equations, presenting the governing equations and detailing numerical implementation, including an efficient autograd scheme leveraging the tridiagonal RANS discretisation, and deriving the adjoint for the ML-augmented $k$--$\omega$ equations. Section 5 presents the numerical results of applying oRANS to turbulent channel flow across a range of $Re_\tau$, where it generally improves mean profiles and turbulent kinetic energy relative to a baseline and offline ML-RANS models, remains stable for modest embedded lengths where full periodic DNS spuriously laminarises, and scales linearly in cost with embedded length. Section~\ref{sec:rotating} applies the method to spanwise-rotating channel flow, a case in which the baseline closure cannot represent the dominant physics, and shows that the online optimisation recovers the rotation-induced asymmetry from non-rotating inflow turbulence and a rotation-blind closure, whereas a model trained offline on the available non-rotating data recovers none of it. Section 7 concludes with a summary and outlook.

The current formulation is limited by the representativeness of the embedded region, boundary-condition contamination, and the under-representation of long-wavelength modes in short domains. These limitations frame the scope of the present work and point toward future extensions, including multi-fidelity RANS/LES solvers with adaptive embedded subdomains.

\section{Online-optimised RANS  (oRANS) algorithm}\label{sec:orans}

The coupled RANS-eDNS system introduced above establishes the basic idea: a high-fidelity subdomain provides reference statistics, while the surrounding RANS domain supplies consistent boundary conditions. We now specialise this framework to the case of conservative PDEs. This class includes most physical systems of interest, and in particular directly covers the three systems studied here: the stochastically forced Burgers equation, incompressible turbulent channel flow and spanwise-rotating channel flow. In this setting, the dynamics are expressed in terms of a state vector, fluxes, sources, and a closure operator, with additional constraints (e.g., continuity) appended where required. This conservative formulation is a concrete realisation of the generic RHS operator $\mathcal{F}$ introduced in section \ref{sec:introduction}, and provides the working form for the adjoint optimisation strategy and algorithmic loop described below. 

\subsection{Governing system and closure}

Consider a general system of nonlinear conservation laws written in conservative form
\begin{equation}\label{eq:HF}
    \frac{\partial \mathbf{Q}}{\partial t} 
    + \nabla \cdot \big(\mathbf{F}(\mathbf{Q}) - \mathbf{F_v}(\mathbf{Q},\nabla \mathbf{Q})\big) 
    = \mathbf{S}(\mathbf{Q}), 
    \quad x \in \Omega,
\end{equation}
where $\mathbf{Q}(x,t)$ is the state vector, $\mathbf{F}$ the inviscid flux, $\mathbf{F_v}$ the viscous flux, and $\mathbf{S}$ source terms. Many systems also impose constraints, such as incompressibility $\mathcal{C}(\mathbf{Q})=\nabla\!\cdot\mathbf{u}=0$, with pressure acting as a Lagrange multiplier. The approach presented below can be easily extended to such incompressible flows which include an additional continuity equation.

Upon averaging or coarse-graining (Reynolds or spatial averaging), unclosed terms appear. The low-fidelity formulation is the conservative analogue of the averaged system of section \ref{sec:introduction},
\begin{equation}\label{eq:LF}
    \frac{\partial \bar{\mathbf{Q}}}{\partial t} 
    + \nabla \cdot \big(\bar{\mathbf{F}}(\bar{\mathbf{Q}}) - {\mathbf{\bar F_v}}(\bar{\mathbf{Q}},\nabla \bar{\mathbf{Q}})\big) 
    = \bar{\mathbf{S}}(\bar{\mathbf{Q}}) 
    + \nabla \cdot h(\bar{\mathbf{Q}};\theta),
\end{equation}
where $h(\cdot;\theta)$ is a closure operator parameterised by $\theta$ (e.g.\ classical coefficients or neural-network weights). In practice, the low-fidelity system $\eqref{eq:LF}$ is discretised in space and time, yielding a nonlinear residual system $\mathbf{R}(\bar{\mathbf{Q}},\theta)=0$. The high-fidelity equations \eqref{eq:HF} are also discretised for simulation but are used solely to generate reference data and do not enter $\mathbf{R}$.

The simulation domain $\Omega$ is partitioned into a high-fidelity embedded subdomain $\Omega_{\mathrm{eDNS}}$ and a low-fidelity subdomain $\Omega_{\mathrm{RANS}}$ with interface
\[
\Gamma = \partial \Omega_{\mathrm{eDNS}} \cap \partial \Omega_{\mathrm{RANS}}.
\]
On $\Omega_{\mathrm{eDNS}}$, the unclosed system~\eqref{eq:HF} is solved directly; on $\Omega_{\mathrm{RANS}}$, the closed system~\eqref{eq:LF} is solved. 
At the interface, the fields are coupled via a transfer operator
\begin{equation}\label{eq:interface}
    \mathbf{Q}|_\Gamma = 
    \mathcal{T}\!\left(\bar{\mathbf{Q}}|_\Gamma;\theta\right),
\end{equation}
which supplies consistent boundary data to the high-fidelity embedded subdomain. 
Conversely, statistics of $\mathbf{Q}$ may feed back into the low-fidelity closure parameters through $\theta$, establishing a two-way coupling. 
In oRANS, $\mathcal{T}$ augments the low-fidelity mean field with rescaled fluctuations to provide statistically representative inflow. (Details for channel flow are given in section~\ref{sec:oranssetup}, and for rotating channel flow in section~\ref{sec:rot_setup}.)

\subsection{Deep neural parameterisation of the closure}

In this work, the closure operator $h(\mathbf{\bar{Q}};\theta)$ is parametrised through a neural network $f_\theta$. Concretely, the network maps local flow features $z$ (e.g. $\mathbf{\bar Q}, \nabla \mathbf{\bar Q}, \nabla^2 \mathbf{\bar Q}$) to a set of effective closure parameters, which are then used to evaluate $h(\mathbf{\bar Q};\theta) = h(\mathbf{\bar Q};f_\theta(z))$. The architecture is designed to capture the strong nonlinear couplings and stiff source terms characteristic of turbulence 
closures. It consists of five hidden layers with two gated residual connections, defined recursively as
\begin{align}
    H^1 &= \sigma(W^1 z + b^1), \nonumber \\
    H^2 &= \sigma(W^2 H^1 + b^2), \nonumber \\
    H^3 &= G^1 \odot H^2, \quad G^1 = \sigma(W^5 z + b^5), \nonumber \\
    H^4 &= \sigma(W^3 H^3 + b^3), \nonumber \\
    H^5 &= G^2 \odot H^4, \quad G^2 = \sigma(W^6 z + b^6), \nonumber \\
    f_\theta(z) &= W^4 H^5 + b^4,
\end{align}
where $\odot$ denotes the Hadamard product, $\sigma$ is a hyperbolic tangent activation for physical smoothness and bounded output, the parameters $\theta$ are the weights $W^k$ and biases $b^k$ of the neural network, and the gate layers $G^1$, $G^2$ are used to allow for modeling the strong nonlinearities expected of fluid turbulence models. We use a constant learning rate of \( \alpha = 10^{-4} \) initially, followed by geometric decay to improve stability. Gradient updates are computed using RMSProp with zero momentum. We observe that model performance is not strongly sensitive to hyperparameter choices, provided sufficient averaging is maintained.

\subsection{Objective functional and adjoint-based optimisation}

The closure parameters $\theta$ are optimised by minimising a mismatch between high and low-fidelity quantities over the embedded domain:
\begin{equation}\label{eq:orans_objective}
    J(\mathbf{\bar Q}) 
    = \int_0^T \int_{\Omega_{\mathrm{eDNS}}} 
    \mathcal{M}\!\left(\bar{\mathbf{Q}}(\theta), \mathbf{Q}\right) \, dx \, dt,
\end{equation}
where $\mathcal{M}$ is a user-defined discrepancy. For the examples presented herein, we minimise the weighted square error in first- and second-order moments
\begin{equation}\label{eq:M}
    \mathcal{M}(\overline{\mathbf{Q}},\mathbf{Q}) = \frac{1}{2}\left(\left|\left|\overline{u^{\mathrm{eDNS}}} - \bar u^\mathrm{RANS}\right|\right|_2^2 + w_k\left|\left|\overline{k^\mathrm{eDNS}} - k^\mathrm{RANS}\right|\right|_2^2\right).
\end{equation}
Note that all dependence of $J$ on $\theta$ is through the state variables $\bar{\mathbf Q}(\theta)$.
After spatial and temporal discretisation, the low-fidelity system yields nonlinear residual equations $\mathbf{R}(\bar{\mathbf{Q}},\theta) = 0$, leading to the optimisation problem
\begin{equation}
    \min_\theta J(\bar{\mathbf{Q}},\theta)
    \quad \text{s.t.} \quad \mathbf{R}(\bar{\mathbf{Q}},\theta) = 0.
\end{equation}
We form the discrete Lagrangian
\begin{equation}
    \mathcal{L}(\bar{\mathbf{Q}},\theta,\hat{\mathbf{Q}}) 
    = J(\bar{\mathbf{Q}}) - \hat{\mathbf{Q}}^\top \mathbf{R}(\bar{\mathbf{Q}},\theta),
\end{equation}
with adjoint variables $\hat{\mathbf{Q}}$. Here $\nabla_\theta$ denotes the total derivative with respect to parameters $\theta$, while $\partial/\partial(\cdot)$ denotes partial derivatives holding other arguments fixed.
Differentiating $\mathcal{L}$ with respect to $\theta$ along feasible trajectories (i.e., a solution $\bar{\mathbf{Q}}$ which satisfies $\mathbf{R}(\bar{\mathbf{Q}},\theta) = 0$) gives
\begin{equation}
    \nabla_\theta \mathcal{L}
=  \left(\frac{\partial J}{\partial \bar{\mathbf{Q}}} - \hat{\mathbf{Q}}^\top \frac{\partial \mathbf{R}}{\partial \bar{\mathbf{Q}}}\right)\frac{ d \bar{\mathbf{Q}}}{ d \theta} - 
\hat{\mathbf{Q}}^\top \frac{\partial \mathbf{R}}{\partial \theta}.
\end{equation}
Eliminating the computationally expensive Jacobian $\tfrac{d\overline{\mathbf{Q}}}{d\theta}$ yields the discrete adjoint equations
\begin{equation}\label{eq:adjoint}
    \left(\frac{\partial \mathbf{R}}{\partial \mathbf{\bar Q} } \right)^\top \mathbf{\hat Q} = \left(\frac{\partial J}{\partial \mathbf{\bar Q}}\right)^\top,
\end{equation}
where $\partial \mathbf{R}/\partial \bar{\mathbf{Q}}$ is the Jacobian of the nonlinear residual evaluated at the forward solution. Although the forward PDE solve is nonlinear, its adjoint is always linear, which enables the use of efficient linear solvers. Moreover, since the same Jacobian appears in the Newton iterations of the forward problem, the adjoint system can reuse the existing forward linear algebra infrastructure.

At feasible points where $\mathbf{R}=0$, $\nabla_\theta \mathcal{L} = \nabla_\theta J $. Therefore, the objective function gradient can be efficiently evaluated via
\begin{equation}\label{eq:adjoint_grad}
 \nabla_\theta J =  - \mathbf{\hat Q}^\top \frac{\partial \mathbf{R}}{\partial \theta}.
\end{equation}
Crucially, we do not differentiate through the high-fidelity solution; the high-fidelity fields enter $J$ as fixed reference data.
Rather than construct adjoint PDEs explicitly, we apply reverse-mode automatic differentiation to the scalar auxiliary function 
\begin{equation}
    \Psi(\bar{\mathbf{Q}};\theta) = \hat{\mathbf{Q}}^\top \mathbf{R}(\bar{\mathbf{Q}},\theta),
\end{equation}
to construct the adjoint equation, treating $\hat{\mathbf{Q}}$ as fixed coefficients. Differentiation of the above scalar auxiliary function with respect to $\bar{\mathbf{Q}}$ reproduces the left-hand side of the discrete adjoint system~\eqref{eq:adjoint}, while differentiation with respect to $\theta$ yields the gradient of the objective function via equation~\eqref{eq:adjoint_grad}.

\subsection{oRANS implementation}\label{sec:implementation}

We discretise time with a fine grid $\{t_n\}_{n\ge0}$ for PDE integration and a coarser sequence of parameter update times $\{\tau_m\}_{m\ge0}$ with $\tau_m=t_{n_m}$ and $n_{m+1}-n_m=M$ (e.g.\ $M=100$). The parameters $\theta$ are held fixed between updates:

\begin{enumerate}
\item Initialise $\mathbf{Q}$ on $\Omega_{\mathrm{eDNS}}$, $\bar{\mathbf{Q}}_{\theta_0}$ on $\Omega_{\mathrm{RANS}}$, and set $\theta_0$.
\item For $m=0,1,2,\dots$ until convergence:
  \begin{enumerate}
    \item Forward PDE solve (fine loop): For $n=n_m,\dots,n_{m+1}-1$, advance the coupled high/low-fidelity system~\eqref{eq:HF}-\eqref{eq:interface} from $t_n$ to $t_{n+1}$ with $\theta=\theta_m$ fixed, enforcing $\mathbf{Q}|_\Gamma=\mathcal{T}(\bar{\mathbf{Q}}|_\Gamma;\theta_m,t)$ at each step.
    \item Adjoint solve (coarse step): At $t=\tau_{m+1}$, form and solve the nonlinear low-fidelity residual system $\mathbf{R}(\bar{\mathbf{Q}}_{\theta_m})=0$
    to obtain the state $\bar{\mathbf{Q}}_{\theta_m}$. The adjoint variables are then computed by solving the linear system in equation \ref{eq:adjoint}.
    For steady low-fidelity systems (e.g.\ RANS or time-averaged Burgers), this adjoint is steady; for unsteady low-fidelity systems it is integrated backward over $[t_{n_m},t_{n_{m+1}}]$.
    \item Parameter update:
    The adjoint solution provides the gradient of the objective with respect to the closure parameters through equation \ref{eq:adjoint_grad},
    avoiding any need to compute $\nabla_\theta \bar{\mathbf{Q}}$. A gradient-descent step is hence applied:
    \begin{equation}
      \theta_{m+1}
      = \theta_{m} - \alpha_m
        \int_{\tau_m}^{\tau_{m+1}}
        \int_{\Omega_{\mathrm{eDNS}}}
        \nabla_\theta J\big(\bar{\mathbf{Q}}_{\theta_m},\mathbf{Q}\big) \, dx \, dt,
      \label{eq:param-update-generic}
    \end{equation}
    with learning rate $\alpha_m$. Note that for quadratic choices of $\mathcal{M}$ including equation \ref{eq:M}, the integrand reduces to the familiar form $(\bar{\mathbf{Q}}-\mathbf{Q})\,\nabla_\theta \bar{\mathbf{Q}}$.
    \end{enumerate}
\end{enumerate}

\noindent The key feature is its online nature: closure parameters are updated concurrently with PDE integration, in contrast to offline regression against precomputed datasets. This conservative-form specialisation of the generic framework in section \ref{sec:introduction} underpins the specific implementations in section \ref{sec:burgers} (Burgers), section \ref{sec:nsresults} (channel flow), and section \ref{sec:rotating} (rotating channel flow).

\section{Validation on Burgers equation}\label{sec:burgers}

To verify the oRANS optimisation mechanics and evaluate its performance in a controlled setting, we first consider the stochastically forced, one-dimensional viscous Burgers equation. This canonical test case retains essential mathematical features of the  Navier--Stokes turbulence cascade, including nonlinear advective and dissipative dynamics, while permitting detailed analysis and rapid numerical experimentation. 

\subsection{Governing equations}

We take the high-fidelity system as the stochastically forced, one-dimensional viscous Burgers equation, a specialisation of~\eqref{eq:HF}. The state, fluxes, and source are
\begin{align}
    \mathbf{Q} = u(x,t), \qquad
    \mathbf{F} &= \frac{1}{2}u^2, \qquad
    \mathbf{F_v}= \frac{1}{\mathrm{Re}}\,\frac{\partial u}{\partial x}, \nonumber \\
    \mathbf{S} &= f_\text{det}(x,t) + \varphi_0 \varphi(x,t), \label{eq:burgersnondim}
\end{align}
where $\mathrm{Re}$ is the Reynolds number, $f_\text{det}$ is a deterministic forcing, and $\varphi(x,t)$ is a unit-variance stochastic process
\citep{Chambers1987, Chambers1988}. The coefficient $\varphi_0$ sets the
forcing amplitude.

Applying Reynolds decomposition $u = \bar u + u'$ to the stochastically forced system yields the low-fidelity equation for the mean state
\begin{equation}
    \frac{\partial \bar u}{\partial t}
    + \bar u \frac{\partial \bar u}{\partial x}
    = \frac{1}{\mathrm{Re}} \frac{\partial^2 \bar u}{\partial x^2}
    + \bar f_\text{det}(x,t)
    + \nabla \cdot h(\bar u;\theta),
\end{equation}
where the closure term $h(\bar u;\theta)$ represents the effect of the
unclosed correlation $\tfrac{1}{2} \overline{u'u'}$.  

\subsection{Burgers equation closure modelling}

The closure term can, in principle, be entirely represented by a neural network or a simple eddy-viscosity closure, for example the zero-equation mixing-length model $\nu_t=C_\mu \ell_m(\theta)^2\left|\frac{\partial \bar{u}}{\partial x}\right|$. Such a model can have large degrees of freedom but may not generalise well. Instead, we derive a single-equation turbulence model in the spirit of Boussinesq-type RANS closures and introduce an augmented low-fidelity state including the ``turbulent kinetic energy'' $k=\tfrac{1}{2}\overline{u' u'}$, $\bar{\mathbf{Q}} = \{\bar u, k\}$. 

In one dimension, assuming the Kolmogorov hypothesis, the Reynolds-stress term is modelled as
\begin{equation}
    \overline{u'u'} = 2 C_\mu k^{1/2}\ell_m \frac{\partial \bar u}{\partial x}.
\end{equation}
The turbulent kinetic energy then evolves according to
\begin{equation}
    \frac{\partial k}{\partial t} + \bar u \frac{\partial k}{\partial x}
    = \frac{\partial}{\partial x}\!\left(\nu_t \frac{\partial k}{\partial x}\right)
      - \frac{1}{Re} C_D \frac{k^{3/2}}{\ell_m}
      + 2 \nu_t \left(\frac{\partial \bar u}{\partial x}\right)^2,
\end{equation}
with $\nu_t = C_\mu k^{1/2}\ell_m$. The reference values $C_\mu=0.55$ and $C_D=C_\mu^3$ used below are the standard one-equation-model constants \citep[\S10.3]{pope2000}. The dissipation term $C_D k^{3/2}/\ell_m$ is the standard one-equation ($k$--$\ell$) model of the dissipation rate, in which the dissipation scales with the turbulence velocity $k^{1/2}$ and the mixing length \citep[see][\S10.3]{pope2000}. The coefficient $C_D$ is one of the three closure parameters learned below. In the conservative notation of section \ref{sec:orans}, the low-fidelity system can thus be written as
\begin{align}
    \bar{\mathbf{Q}} = \left[ \bar u, k\right]^T, \qquad \mathbf{F}&=\left[\frac{1}{2}\bar u^2, \bar u k\right]^T, \qquad \mathbf{F_v}=\left[\frac{1}{Re}\frac{\partial \bar u}{\partial x} - k,\ \nu_t \frac{\partial k}{\partial x}\right]^T, \nonumber \\
    \mathbf{S}&=\left[\bar f_{\mathrm{det}}, -\frac{1}{Re}C_D\frac{k^{3/2}}{\ell_m} +2\nu_t \frac{\partial \bar u}{\partial x} \frac{\partial \bar u}{\partial x} + k\frac{\partial u}{\partial x}\right]^T, \label{eq:burgersrans}
\end{align}
and contains three turbulence parameters: $C_D$, $C_\mu$, and $\ell_m$. These are represented through the closure map
\begin{equation}
    C_D,\, C_\mu,\, \ell_m
    = f_\theta(\bar{\mathbf{Q}}, \nabla \bar{\mathbf{Q}}, \nabla^2 \bar{\mathbf{Q}};\theta),
\end{equation}
so that $\theta$ parameterises the dependence of the closure coefficients on local mean-flow and $k$ features.

The turbulence model introduced above is not unique, and the oRANS framework allows for flexible balancing of physical modelling assumptions and machine learning closure. As in traditional offline machine learning approaches to turbulence modelling, the selection of a suitable model reflects a trade-off between the number of degrees of freedom and generalisability. However, since oRANS is trained on data from the \textit{in situ} flow under identical boundary and physical conditions, generalisation constraints are relaxed compared to conventional machine learning approaches. This permits the use of more expressive machine learning models than would typically be feasible in offline settings.

\subsection{Specification of the stochastic forcing}

The stochastic forcing $\varphi$ is assumed uncorrelated in space and correlated in time, modelled as a sum of Ornstein-Uhlenbeck-driven Fourier modes,

\begin{equation}
\varphi (x,t) \;=\;
\sum_{k\in\{8,16,24,48\}} X_k(t)\sin(2\pi k x),
\quad
d X_k = -\lambda_k X_k\,d t + \sigma\,dW_k,
\end{equation}
with independent Wiener processes $W_k( t )$, decay rates
$\lambda_k=\{1,2,4,8\}$, and noise amplitude $\sigma=10$.  
The prefactor $\varphi_0$ is then selected so that the space-time variance of the
forcing satisfies $\overline{\varphi^2}=1$:
\begin{equation}
\varphi_0
\;=\;
\Bigl(\tfrac{\sigma^{2}}{4}\sum_{k}\lambda_k^{-1}\Bigr)^{-1/2}.
\end{equation}
We additionally apply a deterministic, spatially periodic body-force
\begin{equation}
    f_{\mathrm{det}}(x) = \frac{\sigma}{\varphi_0}\sin(2\pi\cdot4 x),
\end{equation}
which injects energy at a fixed wavenumber and maintains a statistically stationary mean flow.

\subsection{Burgers equation results}

Equations \ref{eq:burgersnondim} and \ref{eq:burgersrans} are solved for $\varphi_0=0.146$, Re $=1300$ on a periodic domain $0\leq x\leq 2\pi$. The stochastic governing equation is solved on an embedded subdomain, while the remaining region is solved using the averaged Burgers equations. Derivatives are computed across the interface, which facilitates the exchange of momentum and energy fluctuations across the two models. The averaged model is run within the embedded region, using the same parameters, to compute gradient updates via the adjoint. Averages are taken over $M=1000$ independent realisations to provide target data for the closure model. A baseline case with no turbulence model $(\overline{u'u'}=0)$ is included as a benchmark for comparison.
\par
\begin{figure}
    \centering
    \includegraphics[width=\textwidth]{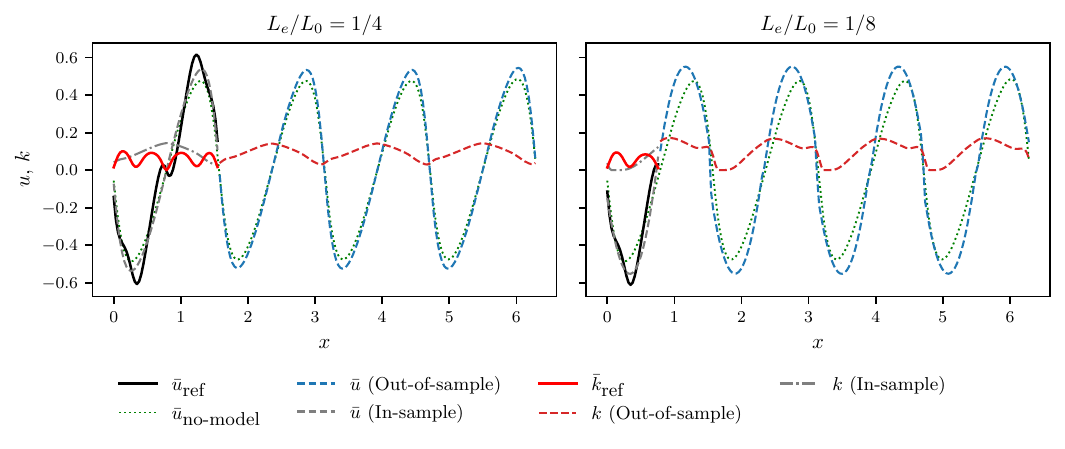}
    \caption{Solutions to the stochastically forced Burgers equation with embedded domains of size (left) $L_e/L_0=1/4$ and (right) $L_e/L_0=1/8$. The panels compare the true means $\bar{u}_\textrm{ref}$,$k_\textrm{ref}$ with online-trained model predictions (in-sample and out-of-sample), and no-model baseline. The results show that the trained model reproduces the true statistics more accurately than the no-model baseline, including in out-of-sample settings.}
    \label{fig:Burgers}
\end{figure}

Figure \ref{fig:Burgers} shows results for two embedded domain sizes: (i) a full period of the longest forcing wavelength and (ii) a half-period, which cannot fully resolve the dominant mode. Here $L_e$ denotes the length of the embedded subdomain and $L_0$ the full periodic domain. In both cases, the online-trained model captures the true statistics more accurately than the baseline.
Quantitative comparisons are provided in table \ref{tab:burgersresults}, where the relative errors are defined as
\begin{align}
    \overline{J_u} &= \frac{\int_\Omega \left(\bar{u}_\textrm{ref}-\bar{u}\right)^2dx}{\int_\Omega \left( \bar{u}_\textrm{ref} - \bar{u}_{\textrm{no-model}}\right)^2dx}, \\
    \overline{J_k} &= \frac{\int_\Omega \left(k_{\textrm{ref}}-k\right)^2 dx}{\int_\Omega \left( k_{\textrm{ref}} - k_{\textrm{no-model}}\right)^2 dx}.
\end{align}
Here $\bar{u}_\textrm{ref}$ and $k_{\textrm{ref}}=\frac{1}{2}\overline{u'u'}$ denote the time-averaged reference profiles obtained from solutions to the stochastic equation \ref{eq:burgersnondim}, $\bar u$ and $k$ are the corresponding predictions from the model under consideration, and $\bar u_{\textrm{no-model}}$, $k_{\textrm{no-model}}$ are the baseline predictions without a closure model.

\begin{table}
\centering
\caption{Relative error in mean velocity and turbulent kinetic energy for the oRANS model, normalised by the no-model baseline. Values below unity indicate improvement, confirming the gains observed in figure~\ref{fig:Burgers} for both embedded domain sizes.}
\label{tab:burgersresults}
\begin{tabular}{ccc}
\toprule
Domain Fraction & $\overline{J_u}$ & $\overline{J_k}$ \\
\midrule
$L_e/L_0=1/4$ & 0.771 & 0.738 \\
$L_e/L_0 = 1/8$ & 0.871 & 0.927 \\
\bottomrule
\end{tabular}
\end{table}

\begin{figure}
    \centering
    \includegraphics[width=0.5\linewidth]{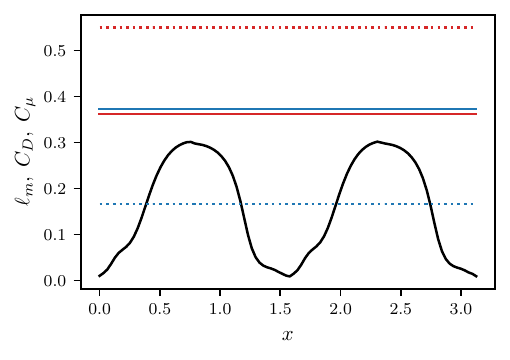}
    \caption{Learned turbulence closure functions for stochastically forced Burgers turbulence. The oRANS-ML trained mixing length $\ell_m^{\mathrm{NN}}$ (black) and turbulence model coefficients $C_\mu^{\mathrm{NN}}$ (red) and $C_D^{\mathrm{NN}}$ (blue) are shown, with standard constants for Navier--Stokes turbulence (Pope, 2000) indicated by dotted lines ($C_\mu=0.55$, $C_D=C_\mu^3$). The learned parameters deviate from the classical values and vary spatially, reflecting adaptation to the flow.
}
    \label{fig:burgers_turbulence}
\end{figure}

The online-trained closure achieves a normalised $L_2$ error reduction of approximately 23\% for velocity and 26\% for turbulent kinetic energy in the full-period case, with moderate degradation in the half-period setup. Figure~\ref{fig:burgers_turbulence} shows the learned mixing length $\ell_m^{\mathrm{NN}}$ and model coefficients $C_\mu^{\mathrm{NN}}$ and $C_D^{\mathrm{NN}}$, and compares to the classical Navier--Stokes constants from \citet{pope2000}. The learned coefficients deviate from the canonical values, and $\ell_m$ adapts to the local forcing scale. Even in the truncated-domain case, the network improves over the baseline by adjusting the production-dissipation balance.

These results indicate that the oRANS framework is robust to moderate under-resolution, and that it adapts effectively to local turbulence characteristics. This motivates future applications to Navier--Stokes flows and more complex geometries.

\section{Navier--Stokes turbulence and closure modelling}

Building on the Burgers verification case, we now move to apply oRANS to Navier--Stokes turbulence. We start by outlining the governing continuum equations alongside numerical implementation with results for the turbulent channel flow deferred to section \ref{sec:nsresults}.

\subsection{DNS governing equations}

The incompressible Navier--Stokes equations are solved in a three-dimensional cuboid subdomain \( \Omega_{\textrm{eDNS}} \subset \mathbb{R}^3 \). In the notation of section~\ref{sec:orans}, the momentum equations are specified by the state, flux, and source vectors
\begin{align}
    \mathbf{Q} &= [u_i], \nonumber \\
    \mathbf{F}(\mathbf{Q}) &= [u_i u_j + p\delta_{ij}\,], \nonumber  \\
    \mathbf{F_v}(\mathbf{Q}, \nabla \mathbf{Q}) &= 
    \left[ \frac{1}{\mathrm{Re}_b}\frac{\partial u_i}{\partial x_j}\,\right], \nonumber  \\
    \mathbf{S}(\mathbf{Q}) &= [f_i\,], \label{eq:momentum}
\end{align}
where \(u_i \in \mathbb{R}^3\) is the velocity, \(p\) the pressure, 
\(\mathrm{Re}_b = u_b (2\delta)/\nu\) the bulk Reynolds number based on bulk velocity \(u_b\), 
kinematic viscosity \(\nu\), and full channel height \(2\delta\), and \(f_i\) is an external forcing. Incompressibility is imposed as the constraint $
\mathcal{C}(\mathbf{Q}) \equiv \nabla \cdot u = 0$, with $p$ acting as a Lagrange multiplier.

Periodic boundary conditions are applied in the streamwise (\(x\)) and spanwise (\(z\)) directions, and no-slip wall conditions in the wall-normal (\(y\)) direction. In cases where fully developed periodic channel flow is not assumed, a Dirichlet inflow and convective outflow condition is imposed in the streamwise direction.

\subsection{DNS numerical implementation}

The governing DNS equations \ref{eq:momentum} are discretised using a second-order central finite difference method on a staggered, structured grid. Velocity components are stored at cell faces, and pressure is located at cell centres. Temporal integration is performed using a classical four-stage Runge--Kutta (RK4) scheme applied to the momentum equations. A fractional-step projection method \citep{Chorin1968} is used to enforce incompressibility, whereby the pressure is computed from a Poisson equation derived by taking the divergence of the momentum equation and applying the continuity constraint:
\begin{equation}
    \nabla^2 p = - \frac{\partial u_i}{\partial x_j} \frac{\partial u_j}{\partial x_i}. \label{eq:poisson}
\end{equation}
The Poisson equation is solved at each Runge--Kutta substep using the BiCGStab iterative solver with a multigrid preconditioner from the \textit{HYPRE} library. This step is parallelised across MPI ranks, while the advection-diffusion updates are advanced using pure MPI halo exchanges. Homogeneous Neumann boundary conditions are applied to the pressure on all boundaries. In this formulation, pressure serves as a Lagrange multiplier that enforces the divergence-free constraint.

To maintain a prescribed bulk Reynolds number $Re_b$, a spatially uniform forcing term $f_i(t) = (f_x(t), 0, 0)$ is applied in the streamwise momentum equation. The magnitude of $f_x(t)$ is updated dynamically at each timestep to enforce
\begin{equation}
    \frac{1}{V}\int_\Omega u(x,y,z,t)dV = 1,
\end{equation}
where $V$ is the volume of the computational domain \citep{Moser1999}. The formulation is equivalent to imposing a time-dependent mean streamwise pressure gradient $-\partial\overline{p}/\partial x=f_x(t)$, and the time-averaged forcing is equal to the mean pressure gradient required to sustain the prescribed bulk velocity (unity in nondimensional units).

The solver has been validated by reproducing the benchmark DNS results of \citet{Kim1987} at \( Re_\tau = 180 \), see \ref{app:validation}, including mean velocity profiles, turbulence intensities, and Reynolds stress distributions. Grid convergence and timestep sensitivity were also verified at this Reynolds number. The simulation parameters for the cases considered herein are shown in table \ref{tab:channelparams}.
\par
Time-averaged flow statistics are computed after the initial transients decay, typically after \( T_{\text{init}}^+ = 500 \). Averaging is performed over an interval of \( T_{\text{avg}}^+ = 5000 \) viscous time units, which corresponds to approximately 50 eddy turnover times. Instantaneous fields are sampled at regular intervals for later analysis. The timestep is chosen to maintain a maximum CFL number below 0.5 in all simulations.

\begin{table}
\centering
\caption{Simulation parameters for reference DNS turbulent channel flow at the friction Reynolds numbers $Re_\tau$ considered. Domain sizes are given in units of the channel half-height $\delta$, and the resolutions ensure grid spacing within DNS standards for near-wall turbulence.}
\label{tab:channelparams}
\begin{tabular}{cccc}
\toprule
\( Re_\tau \) & Domain size \( (L_1, L_2, L_3)/\delta \) & Grid resolution \( (N_1, N_2, N_3) \) & Grid spacing \(\Delta x^+,\Delta y^+_{\min},\Delta z^+ \) \\
\midrule
160 & \( 4\pi \times 2 \times 2\pi \) & \( 192 \times 128 \times 160 \) & $10.5,\ 0.16,\ 6.3$ \\
180 & \( 4\pi \times 2 \times 2\pi \) & \( 192 \times 128 \times 160 \) & $11.8,\ 0.18,\ 7.1$ \\
270 & \( 4\pi \times 2 \times 2\pi \) & \( 384 \times 256 \times 320 \) & $8.8,\ 0.13,\ 5.3$ \\
395 & \( 4\pi \times 2 \times 2\pi \) & \( 384 \times 256 \times 320 \)  & $12.9,\ 0.19,\ 7.76$\\
590 & \( 4\pi \times 2 \times 2\pi \) & \( 768 \times 256 \times 640 \) & $9.7,\ 0.28,\ 5.8$\\
\bottomrule
\end{tabular}
\end{table}

\subsection{RANS governing equations}

In the general $p=3$ case, the incompressible $k$--$\omega$ transport equations for the mean velocity $\overline{u_i}$, turbulent kinetic energy $k$, and specific dissipation rate $\omega$ \citep{Wilcox2008} are solved  for $\mathbf{\bar Q} = \{\bar u_i, \ k, \ \omega \}$ over $\Omega_\textrm{RANS}$,
\begin{align}\label{eq:RANS}
    \frac{\partial }{\partial x_j}\left(\mathbf{F} - \mathbf{F_v}
    \right) &= \mathbf{S},
\end{align}
where the inviscid-flux, viscous-flux, and source-term vectors are given by
\begin{align}
    \mathbf{F} &= \left[\ \bar u_j \bar u_i + \bar p\delta_{ij}, \ \bar u_j k, \ \bar u_j \omega \right]^T, \\
    \mathbf{F_v} &= \left[\frac{1}{\textrm{Re}_b}\frac{\partial \overline{u_i}}{\partial x_j} -  \overline{u_i' u_j'}, \ \left(\frac{1}{\textrm{Re}_b} + \sigma_{k,\theta} \nu_t\right)\frac{\partial k}{\partial x_j}, \  \left(\frac{1}{\textrm{Re}_b} + \sigma_{\omega,\theta}\nu_t\right)\frac{\partial \omega}{\partial x_j} \right]^T, \\
    \mathbf{S} &= \left[0, P_k - \beta^*_\theta \omega k, \ \frac{\gamma_\theta}{\nu_t}P_k - \beta_{0, \theta}\omega^2 \right]^T, \\
    P_k &= 2\nu_t \overline{S}_{ij} \ \overline{S}_{ij}, \\
    \overline{S}_{ij} &= \frac{1}{2}\left(\frac{\partial \overline{u}_i}{\partial x_j} + \frac{\partial \overline{u}_j}{\partial x_i}\right),
\end{align}
and where incompressibility is imposed as the constraint $
\mathcal{C}(\mathbf{\bar Q}) \equiv \partial\bar u_i/\partial x_i = 0$.

The Boussinesq hypothesis closes the RANS equations through the turbulent viscosity $\nu_t$,
\begin{align}
     \overline{u_i' u_j'} &= -2\nu_t\overline{S_{ij}} + \frac{2}{3}k\delta_{ij} \\
    \nu_t &= \alpha_\theta\frac{k}{\omega}.
\end{align}
For channel flow with statistically homogeneous streamwise and transverse directions, the equations reduce considerably as the mean-flow quantities vary only in the wall-normal ($0\leq y \leq L_2$) direction.

The unclosed RANS coefficients \( \sigma_k, \sigma_\omega, \beta^*, \beta_0, \gamma, \alpha \) are modelled as neural networks \( f_\theta \) with flow features as input variables, see section \ref{sec:dlclosure}. These coefficients are optimised online to match high-fidelity statistics using an adjoint-based PDE-constrained optimisation framework.

Throughout, the bulk Reynolds number $Re_b$ (equivalently the bulk velocity and the viscosity) is prescribed and held fixed, and only the closure coefficients are learned. The default $k-\omega$ RANS constants are $\sigma_k = 1/2, \ \sigma_\omega = 1/2, \ \beta^* = 9/100, \ \beta_0 = 3/40, \ \gamma = 5/9$, \ $\alpha=1$. The boundary conditions at the walls are Dirichlet $\overline{u}(0) = \overline{u}(L_2) = 0$, $k(0) = k(L_2) = 10^{-10}$, $\omega(0) = \omega(L_2) = \frac{6\nu}{(3/40)d^2}$, where $d$ is the distance to the nearest wall. 

\subsection{RANS numerical implementation}\label{sec:ransnumerics}

The governing equations \eqref{eq:RANS} are advanced in a fully coupled manner, with the mean-flow and turbulence transport equations solved simultaneously. This monolithic treatment avoids splitting errors and naturally accounts for the coupling between velocity, turbulent kinetic energy, and dissipation rate.

The solution is updated according to a block semi-implicit scheme \citep{Wilcox2006},
\begin{align}\label{eq:RANSdiscretised}
    \left[\frac{I}{\Delta t} + \delta_x\left(\frac{\partial \mathbf{F}}{\partial \mathbf{\bar Q}} - \frac{\partial \mathbf{F_v}}{\partial \mathbf{\bar Q}}\right) - \frac{\partial \mathbf{S}}{\partial \mathbf{\bar Q}}\right]\Delta \mathbf{\bar Q} &= -\delta_x\left(\mathbf{F}^n - \mathbf{F}^n_\mathbf{v} \right) + \mathbf{S}^n.
\end{align}
The source term $\mathbf{S}$ is treated such that $k$ and $\omega$ remain positive semi-definite \citep{Spalart2002} by treating the production terms explicitly and dissipation terms implicitly. Specific to channel flow, where only $\bar u_1$ varies in $y$, the source vector reduces to
\begin{align}\label{eq:RANSdiscretised_S}
    \mathbf{S} = \left\{
    \begin{aligned}
    0 &\\
    \nu_t \frac{\partial \overline{u}}{\partial y}\frac{\partial \overline{u}}{\partial y} &- \beta^*\frac{\omega}{k}k^2 \\
    \gamma\nu_t \frac{\partial \overline{u}}{\partial y}\frac{\partial \overline{u}}{\partial y}\frac{\omega}{k} &- \beta_0\omega^2
    \end{aligned}
    \right\},
\end{align}
and $\frac{\omega}{k}$ and $\nu_t \frac{\partial \overline{u}}{\partial y}\frac{\partial \overline{u}}{\partial y}$ are treated as constants. The model is evaluated with frozen gradients to avoid contaminating the Jacobian structure.

The mean pressure gradient in the RANS solver is dynamically adjusted to maintain a prescribed bulk velocity. The adjustment is implemented as
\begin{equation}\label{eq:forcing}
    \frac{\partial \overline{p}}{\partial x} = -\frac{1}{\Delta t} \left(1 - u_b\right) - \tau_w,
\end{equation}
where \( \tau_w = \nu \left.\frac{\partial \overline{u}_1}{\partial y}\right|_{\text{wall}} \) is the instantaneous wall shear stress. Here the channel half-width and bulk velocity have been normalised to unity. Throughout, the bulk velocity $u_b$ is the volume average of the streamwise velocity over the channel, $u_b=V^{-1}\int_V\bar u_1\,\mathrm{d}V$, which for the RANS solution reduces to the wall-normal average of $\bar u_1$. It is prescribed, and therefore identical for all closures of a given case.

\subsubsection{Efficient Jacobian computation via stencil-wise AD}
To avoid forming dense Jacobians using automatic differentiation, we exploit the block-tridiagonal structure of the semi-implicit discretisation in equations \ref{eq:RANSdiscretised}--\ref{eq:RANSdiscretised_S}. Instead of differentiating the full residual, we apply reverse-mode AD to \textit{localised three-point stencils}, i.e. the per-cell residual contributions, in parallel across the grid via batched Jacobian-vector products. This preserves sparsity and yields the three block diagonals directly, which we assemble into the Newton system and solve with a block-tridiagonal routine. In contrast to standard autograd that materialises a dense Jacobian, our stencil-wise approach reduces peak memory and wall time by more than an order of magnitude on the channel-flow cases reported while retaining the same linear algebra as the forward scheme. The resulting linear solves remain $O(N_2)$ in both time and memory for $N_2$ wall-normal cells, versus $O(N_2^2)$ memory if a dense Jacobian were constructed. This formulation enables efficient and stable implicit time stepping for ML-augmented RANS closures.

The RANS solver is validated against the Wilcox $k$-$\omega$ model (default coefficients) in \textit{OpenFOAM} for turbulent channel flow at \(Re_\tau = 180\), reproducing the benchmark data of \citet{Kim1987}.

\subsection{Adjoint formulation}

We define an objective functional that penalises mismatch in the first- and second-order moments,
\begin{equation}\label{eq:objective_uk}
J(\theta) = \frac{1}{2} \int_{\Omega_{\textrm{eDNS}}} \sum_{i=1}^p\left\| \overline{u}^{\textrm{RANS}}_{i,\theta} - \overline{u}_i^{\textrm{DNS}} \right\|_2^2 + w_k\left\| k^{\textrm{RANS}}_\theta - \overline{k}^{\textrm{DNS}} \right\|_2^2 \, dx,
\end{equation}
where \( \overline{u}_i^{\textrm{DNS}}\) and \(\overline{k}^{\textrm{DNS}} \) are time-averaged quantities obtained from the concurrent embedded simulation, \( \overline{u}^{\textrm{RANS}}_{i,\theta} \) and \( k^\textrm{RANS}_\theta \) are computed from the current RANS model closure defined by \(\theta\), and \(w_k\) weights the two terms. The mean-velocity and kinetic-energy mismatches differ in magnitude by one to two orders, so \(w_k\) is set adaptively. At each update it is moved towards the ratio of the current velocity mismatch to the current kinetic-energy mismatch, so that the two terms contribute equally to the objective, with exponential smoothing between updates and bounds of $10^{-2}\leq w_k\leq10^{2}$.

The governing RANS residuals \eqref{eq:RANS} are written compactly as $\mathbf R(\bar{\mathbf Q},\theta) = 0$, and the discrete Lagrangian then reads
\begin{equation}
    \mathcal{L}(\bar{\mathbf Q},\hat{\mathbf Q},\theta) 
= J(\bar{\mathbf Q},\theta) - \hat{\mathbf Q}^\top \mathbf R(\bar{\mathbf Q},\theta)
\label{eq:lagrangian}
\end{equation}
with adjoint variables $\hat{\mathbf Q}=(\hat u_i,\hat k,\hat\omega)$.
The objective-function gradient is obtained by solving the resulting adjoint equations. To derive these, consider the variation of the Lagrangian
\begin{align}
    \delta \mathcal{L} &= \delta J + \int_\Omega \hat{u}_i \delta f_u^i dx + \int_\Omega \hat{p}\delta f_c dx + \int_\Omega\hat{k}\delta f_k dx+\int_\Omega \hat{\omega} \delta f_\omega dx \nonumber \\& +\int_\Gamma(\hat{u}_i f_u^i+\hat{p}f_c+\hat{k}f_k+\hat{\omega}f_\omega)n_k\delta x_k dx,
\end{align}
where for the $k$--$\omega$ system, the residual vector takes the form $\mathbf{R} = [f_c, f_u^i, f_k, f_\omega]^T$, where $f_c$ is the continuity residual, $f_u^i$ the momentum residuals, $f_k$ the turbulent kinetic energy residual, and $f_\omega$ the specific dissipation residual. $\Omega$, $\Gamma$ define the inner and boundary regions respectively. The first-order variations are then given by
\begin{align}
    \delta f_c &= \frac{\partial \delta \overline{u_i}}{\partial x_i}, \\
    \delta f_u^i &= \delta \overline{u_j}\frac{\partial \overline{u_i}}{\partial x_j} + \overline{u_j}\frac{\partial \delta \overline{u_i}}{\partial x_j} + \frac{\partial \delta p}{\partial x_i} \nonumber \\&- \frac{\partial}{\partial x_j}\left[ \left(\frac{1}{\textrm{Re}_b}+\nu_t \right)\left(\frac{\partial \delta \overline{u_i}}{\partial x_j} + \frac{\partial \delta \overline{u_j}}{\partial x_i} \right) \right]  - \frac{\partial}{\partial x_j}\left[ \delta\nu_t\left(\frac{\partial \overline{u_i}}{\partial x_j} + \frac{\partial \overline{u_j}}{\partial x_i}\right)\right], \\
    \delta f_k &= \delta \overline{u_j}\frac{\partial k}{\partial x_j}
     + \overline{u_j}\frac{\partial \delta k}{\partial x_j} - \frac{\partial}{\partial x_j}\left[\left(\frac{1}{\textrm{Re}_b} + \sigma_{k,\theta} \nu_t\right) \frac{\partial \delta k}{\partial x_j} \right] \nonumber \\& - \frac{\partial}{\partial x_j}\left(\sigma_{k,\theta} \delta \nu_t \frac{\partial k}{\partial x_j}\right)  - \delta P_k + \beta^*_\theta \omega\delta k + \beta^*_\theta k \delta \omega, \\
     \delta f_\omega &= \delta \overline{u_j}\frac{\partial \omega}{\partial x_j}
     + \overline{u_j}\frac{\partial \delta \omega}{\partial x_j} - \frac{\partial}{\partial x_j}\left[\left(\frac{1}{\textrm{Re}_b} + \sigma_{\omega,\theta} \nu_t\right) \frac{\partial \delta \omega}{\partial x_j} \right] \nonumber \\&- \frac{\partial}{\partial x_j}\left(\sigma_{\omega,\theta}\delta \nu_t \frac{\partial \omega}{\partial x_j}\right) - \frac{\gamma_\theta \omega}{\alpha_\theta k}\delta P_k  - \frac{\gamma_\theta}{\alpha_\theta k} P_k \delta{\omega} + \frac{\gamma_\theta \omega}{\alpha_\theta k^2}P_k \delta k + 2\beta_{0,\theta}\omega\delta \omega, \\
     \delta \nu_t &= \alpha_\theta \left(\frac{1}{\omega}\delta k - \frac{k}{\omega^2}\delta \omega\right).
\end{align}
Integrating by parts and collecting like terms for the inner domain $\Omega$ then yields the adjoint equations
\begin{align}\label{eq:RANSadjoint}
\frac{\partial \hat{u}_i}{\partial x_i} &=0, \\
- \overline{u_j}\frac{\partial \hat{u}_i}{\partial x_j} + \frac{\partial \overline{u_j}}{\partial x_i}\hat{u}_j -\frac{\partial}{\partial x_j}
\left[\left(\frac{1}{\textrm{Re}_b}+\nu_t\right)
\left(\frac{\partial \hat{u}_i}{\partial{x_j}}+\frac{\partial \hat{u_j}}{\partial {x_i}}\right)\right] \nonumber \\ + \frac{\partial \hat{p}}{\partial x_i} +\hat{k}\frac{\partial k}{\partial x_i} + \hat{\omega}\frac{\partial\omega}{\partial x_i} + 2\frac{\partial}{\partial x_j}\left(\nu_t\,S_{ij}\,\hat{k} + \gamma_\theta\,S_{ij} \hat{\omega}
\right) &= \frac{\partial J}{\partial \overline{u_i}}, \\
- \overline{u_j}\frac{\partial \hat{k}}{\partial x_j} -\;\frac{\partial}{\partial x_j}\left[
\left(\frac{1}{\textrm{Re}_b}+\sigma_{k,\theta}\nu_t\right)
\frac{\partial \hat{k}}{\partial x_j}\right] - \Bigl(\frac{P_k}{k}-\beta^{*}_\theta\omega\Bigr)\,\hat{k} \nonumber \\+ \frac{1}{\omega}
\Bigl(\frac{\partial \overline{u_i}}{\partial x_j}+\frac{\partial \overline{u_j}}{\partial x_i}\Bigr)
\frac{\partial \hat{u}_i}{\partial x_j} +\frac{\sigma_{k,\theta}}{\omega}\,
\frac{\partial k}{\partial x_j}\,
\frac{\partial \hat{k}}{\partial x_j} +\frac{\sigma_{\omega,\theta}}{\omega}\frac{\partial\omega}{\partial x_j}\,
\frac{\partial\hat{\omega}}{\partial x_j} &= \frac{\partial J}{\partial k}, \\
-\overline{u_j}\frac{\partial \hat{\omega}}{\partial x_j} -\;\frac{\partial}{\partial x_j}\left[
\left(\frac{1}{\textrm{Re}_b}+\sigma_{\omega,\theta}\nu_t\right)
\frac{\partial \hat{\omega}}{\partial x_j}\right] + 2\beta_{0, \theta}\omega\hat{\omega} + \beta^{*}_\theta k\hat{k} \nonumber \\- \frac{k}{\omega^2}
\Bigl(\frac{\partial \overline{u_i}}{\partial x_j}+\frac{\partial \overline{u_j}}{\partial x_i}\Bigr)
\frac{\partial \hat{u}_i}{\partial x_j} -\frac{\sigma_{k,\theta} k}{\omega^2}\frac{\partial k}{\partial x_j}\frac{\partial \hat{k}}{\partial x_j} -\frac{\sigma_{\omega,\theta} k}{\omega^2}\frac{\partial\omega}{\partial x_j}\frac{\partial\hat{\omega}}{\partial x_j} &= \frac{\partial J}{\partial \omega}.
\end{align}
Stationarity in $\bar{\mathbf Q}$ therefore gives the discrete adjoint equations, while differentiation in $\theta$ yields the gradient needed for optimization:
\begin{equation}
\nabla_\theta J = -\hat{\mathbf Q}^\top \frac{\partial \mathbf R}{\partial \theta}.
\end{equation}

In practice, however, we do not explicitly discretise the continuous adjoint PDEs expressed above. As described in section~\ref{sec:orans}, we instead evaluate $\hat{\mathbf Q}^\top \partial\mathbf R(\bar{\mathbf Q},\theta)/\partial\bar{\mathbf{Q}}$ and $\hat{\mathbf Q}^\top \partial\mathbf R(\bar{\mathbf Q},\theta)/\partial\theta$, resulting from differentiating \eqref{eq:lagrangian} with respect to $\theta$, via reverse-mode automatic differentiation applied to a scalar auxiliary function $\Psi = \hat{\mathbf Q}^\top \mathbf R(\bar{\mathbf Q},\theta)$, treating $\hat{\mathbf Q}^\top$ as constant over single adjoint time steps. This allows us to reuse the forward solver infrastructure for the adjoint system and obtain the gradient without forming Jacobians explicitly.

\begin{figure}
    \centering
    \includegraphics[width=0.4\textwidth]{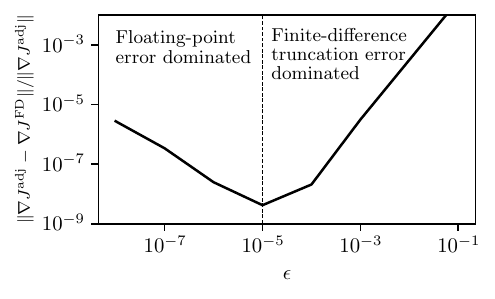}
    \caption{Relative adjoint gradient error as a function of the finite-difference perturbation size $\epsilon$. For small $\epsilon$, floating-point round-off error dominates, while for large $\epsilon$, finite-difference truncation error dominates. This characteristic curve, with low minimum error, confirms the correctness of the adjoint implementation.}
    \label{fig:AdjointVerification}
\end{figure}

To verify the adjoint implementation, we compare adjoint-computed gradients with finite-difference approximations obtained by perturbing the viscosity $1/\textrm{Re}_b$. The finite-difference gradient is computed as $\nabla J^{\mathrm{FD}} = (J(1/Re_b+\epsilon) - J(1/Re_b-\epsilon)) / (2\epsilon)$, where $\epsilon$ is the perturbation size.
Figure~\ref{fig:AdjointVerification} shows the resulting relative gradient error as a function of the finite-difference step size~$\epsilon$. The expected V-shaped curve, with a low minimum error, confirms the correctness of the adjoint formulation.

\subsection{Deep learning closure model}\label{sec:dlclosure}

We now consider applications to the statistically 1D turbulent channel flow case.
We initially consider three different strategies for determining the closure coefficients $\phi_p = \{\alpha, \  \beta^*,\beta_0,\sigma_k,\sigma_\omega,\gamma\}$ or their neural‐network generalisations. In all cases, we denote by $\theta$ the parameters to be optimised.

First, consider a \textit{parametric model}, where the six $k$--$\omega$ closure constants are treated as global scalars i.e.
\begin{equation}\label{eq:parametric}
    \phi_p = \theta_p = \bigl(\alpha, \ \beta^*,\,\beta_0,\,\sigma_k,\,\sigma_\omega,\,\gamma\bigr),
\end{equation}
where $\theta_p \in \mathbb{R}^6$ are directly optimised to minimise the mismatch with DNS data.

Second, consider a \textit{global feature model}, where the closure parameters vary with wall distance, expressed as a function of the non-dimensional coordinate $y^+ = y_\textrm{min} u_{\tau}/\nu$, where $y_{\textrm{min}}$ is the nearest distance to the wall:
\begin{equation}\label{eq:global}
    \phi_g(y) = f_\theta\left(y^+\right),
\end{equation}
where $f_\theta$ is a fully connected network with parameters $\theta$.

Finally, consider a \textit{local flow feature} model, where the closure coefficients depend on local dimensionless flow invariants,

\begin{equation}\label{eq:local}
        \phi_l(y) = f_\theta\left(S^*, Re_T,\left(\frac{\partial k}{\partial y}\right)^+, \left(\frac{\partial \omega}{\partial y}\right)^+\right).
\end{equation}
 The shear rate $S^* = \frac{1}{\omega}\frac{\partial \overline{u}}{\partial y}$ represents the shear to dissipation balance; the turbulent Reynolds number $Re_T = \frac{k}{\nu\omega}$ the turbulence intensity; $\left(\frac{\partial k}{\partial y}\right)^+ = \frac{\partial k}{\partial y} \frac{\nu}{k^{1.5}}$ represents dimensionless turbulent kinetic energy transport and $\left(\frac{\partial \omega}{\partial y}\right)^+ = \frac{\partial \omega}{\partial y}\frac{k^{0.5}}{\omega^2}$ represents the dimensionless turbulent dissipation transport. The network inputs are constructed from dimensionless local invariants, ensuring that the model is scale- and rotation-invariant and generalisable across different Reynolds numbers.
As will be shown in section~\ref{sec:offline}, the local flow feature model $\phi_l$ provides sufficient flexibility to accurately reproduce DNS statistics across a range of Reynolds numbers. Accordingly, we adopt this model exclusively during the online closure optimisation phase.

The inputs to the network are normalised to remain order $O(1)$. 
Each input feature
\begin{equation}
\mathbf{x} = \left[ S^*, \, Re_T, \, \left(\frac{\partial k}{\partial y}\right)^+, \, \left(\frac{\partial \omega}{\partial y}\right)^+\right]^\top,
\end{equation}
is hence divided by a corresponding normalisation coefficient $\mathbf{C}_{\text{Norm}} = \left[ \frac{1}{4}, \, 10, \, 25, \, \frac{1}{10} \right]^\top$ to ensure consistent magnitudes across different quantities,
\begin{equation}
\mathbf{x}_{\text{norm}} = \mathbf{x} \circ \mathbf{C}_{\text{Norm}}^{-1}.
\end{equation}

\subsection{Offline validation of ML closure capacity}\label{sec:offline}

\begin{figure}
    \centering
    \begin{subfigure}{0.48\textwidth}
        \centering
        \includegraphics[width=\textwidth]{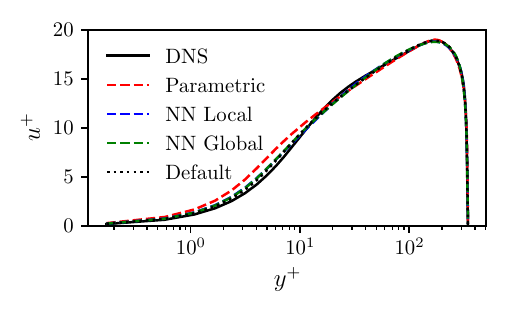}
        \label{fig:u872}
    \end{subfigure}
    \hfill
    \begin{subfigure}{0.48\textwidth}
        \centering
        \includegraphics[width=\textwidth]{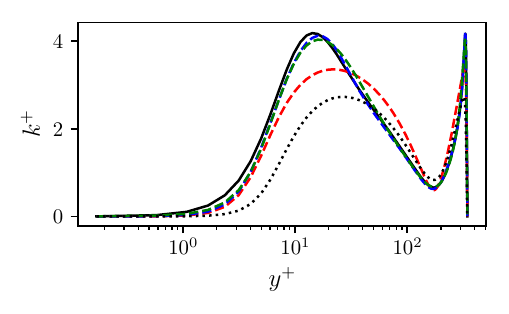}
        \label{fig:k872}
    \end{subfigure}
        \begin{subfigure}{0.48\textwidth}
        \centering
        \includegraphics[width=\textwidth]{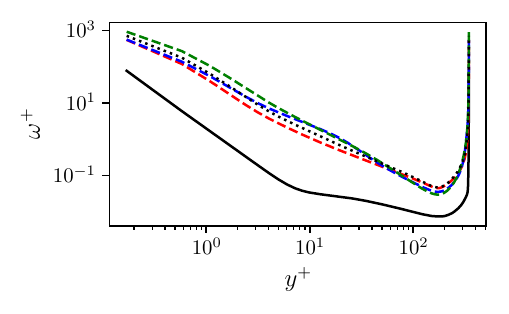}
        \label{fig:w872}
    \end{subfigure}
    \hfill
    \begin{subfigure}{0.48\textwidth}
        \centering
        \includegraphics[width=\textwidth]{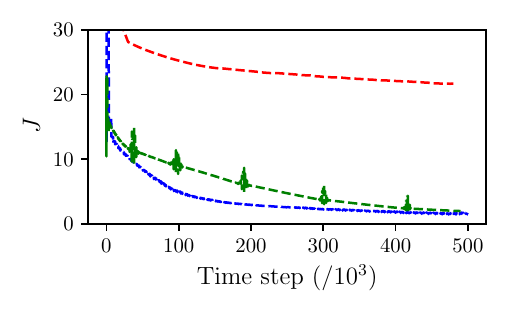}
        \label{fig:J872}
    \end{subfigure}

    \caption{Comparison of RANS closure models at $Re_\tau=180$. Top row: mean velocity $u^+$ (left) and turbulent kinetic energy $k^+$ (right). Bottom row: specific dissipation rate $\omega^+$ (left) and objective function $J$ decay during training (right). Results are shown for DNS, the default $k$--$\omega$ model, a parametric closure (equation~\ref{eq:parametric}), and both local (equation~\ref{eq:local}) and global (equation~\ref{eq:global}) feature neural-network closures. The neural closures closely reproduce DNS statistics and converge effectively, in contrast to the default and parametric baselines; note that $J$ is optimised only over $u$ and $k$, and $\omega^+$ is included only for reference.}\label{fig:offline}
\end{figure}

We first present supervised fits of the DL closure models directly to DNS data for turbulent channel at $Re_\tau=180$, as a demonstration of model flexibility. The training targets are generated \textit{a priori} from long DNS time averages. The results are shown in figure \ref{fig:offline}.

All models, including the default $k-\omega$ model, match the mean velocity profile well. The turbulent kinetic energy is however poorly predicted by the default model, with peak TKE production near half of the DNS prediction, and estimated closer to the centre of the channel.

The parametric model improves on the default model for turbulence statistics predictions, but a more complex model is clearly required to match the DNS data, demonstrating the fundamental limitation of the $k-\omega$ formulation. The NN models are shown to be flexible enough to achieve this, reproducing both velocity and TKE statistics, including the near-wall peak of $k^+$, and convergence is near monotonic and rapid, with little quantifiable difference between local and global models.

The local flow feature neural network (see equation \ref{eq:local}) RANS model is clearly expressive enough to accurately model channel flow turbulence, and relies only on invariant local inputs. We hence adopt the local model architecture for the subsequent online training and deployment within the RANS--embedded-simulation framework, while noting that the present results serve only to demonstrate representational capacity rather than generalisation.

\subsection{oRANS setup for turbulent channel flow}\label{sec:oranssetup}

A note on terminology is in order before the set-up is described. The embedded region $\Omega_{\mathrm{eDNS}}$ is simulated at DNS resolution, but it is short, non-periodic in the streamwise direction, and driven at its inlet by the RANS solution and rescaled fluctuations. Its statistics are therefore conditioned on the model being trained and are not an independent reference in the classical sense. We refer to it as the embedded high-resolution simulation, abbreviated eDNS for its resolution and for consistency with the embedded-DNS literature \citep{He2018,Chen2022}, in which a fine-mesh block at DNS resolution is coupled to a coarse global domain, and reserve the term `reference DNS' for the converged periodic simulations against which all results are compared. DNS resolution is used here so that the method can be assessed against a converged reference. The method does not require it, and in applications to complex systems the embedded region would be resolved by LES (section~\ref{sec:convergence_cost}). The consequences of this conditioning for what the closure can learn are examined directly in section~\ref{sec:rot_length}.

Having established the representational capacity of the neural closure, we next describe its deployment within the oRANS framework for turbulent channel flow. The computational domain $\Omega$ is partitioned into a RANS subdomain $\Omega_{\textrm{RANS}}$ and an embedded region $\Omega_{\textrm{eDNS}}$, such that $\Omega = \Omega_{\textrm{RANS}} \cup \Omega_{\textrm{eDNS}}$. The two solvers exchange boundary conditions across the common interface $\Gamma = \partial \Omega_{\mathrm{RANS}} \cap \partial \Omega_{\mathrm{eDNS}}$.

To benchmark this approach, we consider a target flow at Reynolds number \( Re_2 \) and inject inflow fluctuations derived from a separate fully developed DNS at \( Re_1 \). These are rescaled according to
\begin{align}\label{eq:rescale}
	u^{\textrm{eDNS}}_{\theta} = \overline{u}_{\theta}^{\textrm{RANS}} 
        + \sqrt{\frac{k_{\theta}^{\textrm{RANS}}}{k^\textrm{eDNS}_{Re_1}}}\,
          u^{' \textrm{eDNS}}_{Re_1} := \mathcal{T}, 
        \quad x \in \Gamma,
\end{align}
and used to drive the embedded simulation at \( Re_2 \), while a closure model is concurrently optimised online to reduce the mismatch between embedded and RANS statistics. Although the DNS fields are unsteady, their statistics converge as $t \to \infty$. The oRANS closure therefore adapts to reduce statistical error rather than instantaneous mismatch, enabling generalisation across Reynolds numbers. The solution in $\Omega_{\mathrm{eDNS}}$ depends on $\theta$ through its interface coupling $\mathcal{T}$ with $\Omega_{\mathrm{RANS}}$, and conversely the RANS closure adapts through statistical feedback from $\Omega_{\mathrm{eDNS}}$.

The rescaling adjusts the amplitude of the fluctuations to the RANS statistics but not their structure or spatio-temporal correlations, so the embedded simulation must regenerate whatever structural difference exists between the reference flow and the target flow within the embedded region, and the closure can learn only what is regenerated before the statistics are sampled. Section~\ref{sec:rot_length} measures this adjustment in a flow whose structure differs qualitatively from that of the inflow, and shows both the correction the embedded region performs and the feature it fails to regenerate when too short. In practice, the transformation $\mathcal{T}$ need not be limited to the simple rescaling used here. More sophisticated formulations may be required for complex flows, and a natural extension would be to replace $\mathcal{T}$ with a turbulence-inflow generator consistent with both RANS statistics and local flow features. Moreover, the objective can be augmented to train on additional DNS statistics beyond $(\bar u,k)$; for example, including $\omega$ enables timescale learning and allows $\mathcal{T}$ to be conditioned on $(\bar u, k, \omega)$. By bringing the inlet fluctuations closer to a Navier--Stokes-consistent state, such extensions are expected to reduce boundary-condition contamination and shorten transient adjustment phases. We do not pursue these variants here, but they represent natural directions for extending the framework.

\begin{figure}
    \centering
    \includegraphics{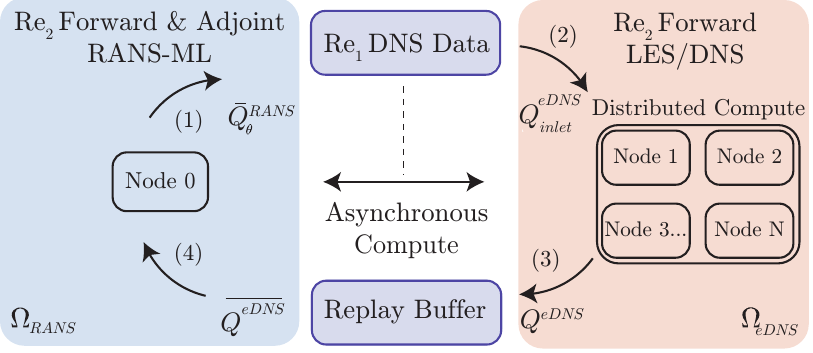}
    \caption{Schematic of the oRANS computational workflow for turbulent channel flow. The RANS-ML solver (blue, $\Omega_{\mathrm{RANS}}$) runs concurrently with the embedded simulation (red, $\Omega_{\mathrm{eDNS}}$), exchanging boundary conditions and statistical feedback through a replay buffer.}\label{fig:schematic_comp}
\end{figure}

The overall computational workflow is shown in figure \ref{fig:schematic_comp}. The RANS-ML solver (Node 0) computes mean fields $\overline{\mathbf{Q}}^{\mathrm{RANS}}_\theta$ and adjoint sensitivities. These statistics rescale stored inflow fluctuations from $Re_1$ DNS, providing inlet conditions to the embedded simulation at $Re_2$ via $\mathcal{T}$. The embedded simulation is run in parallel across multiple nodes, producing high-fidelity fields $\mathbf{Q}^{\mathrm{eDNS}}$ that are accumulated in a replay buffer. Time-averaged statistics from this buffer define the mismatch with the RANS solution, yielding sensitivities that drive the update of $\theta$. The full training procedure is detailed in algorithm \ref{alg:online_optim}.

\begin{algorithm}
\caption{Online Training Algorithm for the RANS closure with an embedded simulation}
\label{alg:online_optim}
\DontPrintSemicolon
\KwIn{DNS data at \( Re_1 \), target Reynolds number \( Re_2 \), learning rate \( \alpha_m \)}
\KwOut{Trained closure parameters \( \theta \)}

\textbf{Time grids:} define the fine grid \( \{t_n\}_{n\ge0} \) with \( t_n = n\Delta \).
Define coarse update times \( \{\tau_m\}_{m\ge0} \) by \( \tau_m = t_{n_m} \) with \( n_{m+1}-n_m = M \).

\begin{enumerate}
  \item Generate turbulent fluctuations from a periodic DNS at \( Re_1 \)
  \item Initialise a replay buffer \(\mathcal{R}\) using one flowthrough of eDNS data at \(Re_2\) seeded with the \(Re_1\) fluctuations.
  
  \item Pretrain closure model parameters \( \theta_0 \) on baseline RANS \( k-\omega \) fields
\item \textbf{For each parameter update time \(\tau_m\), with \(m=0,1,2,\dots\):}
  \begin{enumerate}
    \item \textbf{DNS stepping (fine loop):} For \(n=n_m,\dots,n_{m+1}-1\), run eDNS at \(Re_2\) with inflow
    \[
      u'_{\mathrm{inlet}}(t_n)
      = \sqrt{\tfrac{k^{\mathrm{RANS}}}{k^{\mathrm{eDNS}}}}\,
        u'^{\mathrm{eDNS}}(t_n) + \overline{u}^{\mathrm{RANS}} ,
    \]
    and append downstream statistics \(u^{\mathrm{eDNS}}(t_n)\) to the replay buffer \(\mathcal{R} = \big[u^{\mathrm{eDNS}}(t_{n_m-K}),\;\dots,\;u^{\mathrm{eDNS}}(t_{n_m})\big]\)
    \item Remove old samples from \(\mathcal{R}\) with times \(t<t_{n_m-K}\).
    \item Randomly sample mini-batches \(v \subset \mathcal{R}\) and compute
    time-averaged statistics \(\overline{u}_{\mathrm{DNS}}, \overline{k}_{\mathrm{DNS}}\).
    \item \textbf{Parameter update:}
    \begin{equation}
      \theta_{m+1} = \theta_{m} + \alpha_{m}
      \int_{\Omega_{\mathrm{eDNS}}}\!
      (v - \overline{u}^{\mathrm{RANS}}_{\theta_m})\,
      \nabla_\theta \overline{u}^{\mathrm{RANS}}_{\theta_m}\, dx.
    \end{equation}
\item Apply RMSProp with learning rate \(\alpha_m\), asymptotically minimising
\begin{align}
J(\theta)
= \int_{\Omega_{\mathrm{eDNS}}}
\Bigg(
\Big\|
\lim_{T\to\infty}\frac{1}{T}\int_0^T u^{\mathrm{eDNS}}_\theta(t,x)\,dt
- \overline{u}^{\mathrm{RANS}}_\theta(x)
\Big\|_2^{2}
\;\\+ \ 
w_k\Big(
\lim_{T\to\infty}\frac{1}{T}\int_0^T k^{\mathrm{eDNS}}_\theta(t,x)\,dt
- k^{\mathrm{RANS}}_\theta(x)
\Big)^{2}
\Bigg)\,dx,
\end{align}
where $\overline{u}^{\mathrm{RANS}}_\theta$, $k^{\mathrm{RANS}}_\theta$ are solved on $\Omega_{\mathrm{eDNS}}$.
\item Recompute the RANS solution with updated \(\theta_{m+1}\).
  \end{enumerate}
\end{enumerate}
\end{algorithm}

Online training inside a time-marching solver can fail in two ways. The Newton iteration for the RANS system may fail to converge once the closure has drifted to an unphysical coefficient field, and a single large gradient from a noisy sampled target may move the closure there in one update. The implementation therefore includes the following measures. After each Newton update $k$ and $\omega$ are clipped from below at $10^{-10}$. The Newton iteration is limited to $5000$ iterations, and the parameter update is skipped if the largest Newton increment has not fallen below $10^{-4}$ within this limit, or if the gradient is not finite. The norm of the gradient is clipped and the sampled statistics and $w_k$ are smoothed between updates by an exponential moving average. These measures keep the coupled system bounded when the closure is in error. They cannot correct a biased sampled target, which is the failure mode of the shortest embedded regions (sections~\ref{sec:subdomain_length} and~\ref{sec:rot_length}).

Every RANS result reported below (tables~\ref{tab:wmse_results} and~\ref{tab:rot_jstar}, figure~\ref{fig:Online}) is the solution of the RANS system alone with the learned closure. The embedded simulation generates the training data and does not enter the reported solution.

\section{Turbulent channel flow numerical results}\label{sec:nsresults}

With solvers and closures defined, we now present numerical experiments on turbulent channel flow. We begin with a summary of the main findings, before detailing the results and analysis in section \ref{sec:periodiceDNS} onward. Across all tested Reynolds numbers, oRANS achieves lower errors than the baseline $k$-$\omega$ closure and, in most cases, than offline DL closures. Crucially, it maintains accuracy with modest embedded regions ($L_x \approx 2\pi\delta$), whereas DNS in shortened periodic boxes spuriously laminarises and produces qualitatively incorrect profiles. Because only the embedded region is resolved at high fidelity, the cost of oRANS scales approximately linearly with embedded length. At the same time, the results highlight key limitations that will guide future development. Performance degrades when boundary-condition contamination reaches the interior of short subdomains or when low-wavenumber modes are under-represented.

\subsection{Shortened streamwise domain periodic DNS}\label{sec:periodiceDNS}

As a preliminary diagnostic, we examine the behaviour of DNS in shortened periodic domains.

\begin{figure}
    \centering
    \begin{subfigure}[t]{0.48\linewidth}
        \centering
        \includegraphics[width=\linewidth]{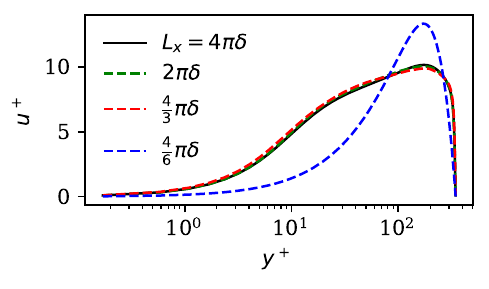}
        \label{fig:u872_2}
    \end{subfigure}
    \hfill
    \begin{subfigure}[t]{0.48\linewidth}
        \centering
        \includegraphics[width=\linewidth]{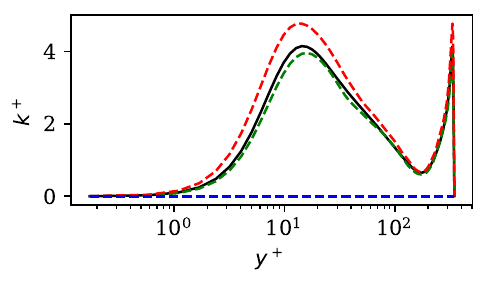}
        \label{fig:k872_2}
    \end{subfigure}
    \caption{Comparison of velocity and turbulent kinetic energy profiles from DNS with periodic boundary conditions at $Re_\tau=180$, using various shortened streamwise domain lengths $L_x$. While domains of length $L_x \geq 2\pi\delta$ reproduce the expected turbulent statistics, excessively short domains (e.g. $L_x=\tfrac{2}{3}\,\pi\delta$) lead to laminarisation.}
    \label{fig:periodicshort}
\end{figure}

This analysis provides guidance for the oRANS framework: the minimum length of the embedded subdomain $\Omega_{\mathrm{eDNS}}$ must be large enough to sustain realistic turbulence. When the streamwise extent of a periodic DNS box is reduced below this threshold, the turbulence dynamics become distorted and in some cases the flow re-laminarises altogether. Figure \ref{fig:periodicshort} shows velocity and TKE profiles at $Re_\tau=180$, where laminarisation is observed for very short boxes (e.g. $L_x=\tfrac{2}{3}\pi\delta$). At higher Reynolds numbers in the present study, laminarisation was not observed; however, the outcome may depend on the initial condition and the basin of attraction of the turbulent state, an issue we regard as beyond the present scope. 

\begin{table}
\centering
\begin{tabular}{lcccccc}
\toprule
Case ($Re_\tau$) & $k$--$\omega$ & \multicolumn{5}{c}{Periodic DNS, $L_x=$} \\
& & $4\pi\delta$ & $2\pi\delta$ & $\tfrac{4}{3}\pi\delta$ & $\tfrac{2}{3}\pi\delta$ & $\tfrac{1}{2}\pi\delta$ \\
\midrule
II (180)  & 1.0 & 0.00 & 0.10 & 0.20 & 131 & 131 \\
III (270) & 1.0 & 0.00 & 0.06 & 0.07 & 0.11 & 0.38 \\
IV (395)  & 1.0 & 0.00 & 0.10 & 0.03 & 0.21 & 0.19 \\
V (590)   & 1.0 & 0.00 & 0.89 & 0.28 & 0.63 & 0.61 \\
\bottomrule
\end{tabular}
\caption{Normalised cost functional $J^*$ of velocity and turbulent kinetic energy for baseline RANS ($k-\omega$), and periodic boundary condition DNS with different streamwise domain fractions $L_x$. Values below $1$ indicate improvement over the baseline RANS. Very large values correspond to laminarisation.}
\label{tab:periodic_results}
\end{table}

To quantify the departure from reference DNS statistics, we introduce a normalised cost functional. For each quantity $q \in \{ u,k\}$ we define
\begin{equation}\label{eq:costfunctional}
J_q = \tfrac{1}{2}\int_{-\delta}^{\delta} \left(q(y) - q^{\mathrm{DNS}}(y)\right)^2 dy,
\end{equation}
and combine them as
\begin{equation}
J^* = \frac{1}{1+w_k}\left(\frac{J_u}{J_{u0}} + w_k \frac{J_k}{J_{k0}}\right), \label{eq:jstar}
\end{equation}
where $w_k$ is set to $5$ and $J_{u0}, J_{k0}$ are the values of $J_u$ and $J_k$ for the baseline $k$--$\omega$ model. By construction, $J^* = 1$ corresponds to the unmodified $k$--$\omega$ model, values below $1$ indicate improvement, and very large values correspond to laminarisation.

Table \ref{tab:periodic_results} shows that the predicted flow statistics deteriorate significantly when the DNS domain length is reduced below approximately \( L_{x,0}/3 \), with very large values corresponding to laminarisation. For domains that remain turbulent, the distortion of statistics manifests primarily as an overprediction of turbulent kinetic energy. This is evident at $L_x=\tfrac{4}{3}\,\pi\delta$ in figure \ref{fig:periodicshort}, where the mean velocity is well captured but the TKE exhibits a clear overshoot, a representative pathology of shortened-domain simulations more generally.

This behaviour is consistent with the concept of a minimal flow unit: for \( Re_\tau \approx 180 \), sustaining near-wall turbulence requires a streamwise extent of at least \( L_x^+ \geq 300 \) \citep{Jimenez1991}, which corresponds to approximately \( L_{x,0}/8 \) in physical space. Below this threshold, the turbulence regeneration cycle is suppressed and the flow reverts to a laminar state. At higher Reynolds numbers the required physical length increases, reflecting the growth of outer-layer structures even as near-wall structures remain of fixed size in wall units.

\subsection{oRANS channel flow experiments}\label{sec:supervisedlearning}

We now present the results following the oRANS approach detailed in algorithm \ref{alg:online_optim}. We summarise the numerical studies considered herein in table \ref{tab:onlineparams}. To place oRANS in context, we compare against three references of increasing fidelity: (i) fully periodic channel-flow DNS with shortened streamwise domains, providing a high-fidelity reduced-cost reference, (ii) a rescaled offline-trained $k$--$\omega$ RANS without online adaptation, representing the standard offline ML strategy, and (iii) state-of-the-art turbulence inflow generation methods \citep{Dreze2023}, which provide synthetic turbulence fluctuations at the inlet based on target statistics.

We simulate a full periodic DNS at a reference Reynolds number to generate turbulent statistics. We then re-scale the statistics based on the online trained RANS solution to the target Reynolds number across a range of channel lengths $L_{x,0}/8\leq L_x\leq L_{x,0}$. Of particular interest is how the statistics degrade for shorter channel lengths, and how rapidly the statistics converge downstream of the inlet.

\begin{table}
\centering
\begin{tabular}{cccc}
\toprule
Case & $Re_b$ of inflow DNS & Target $Re_b$ & Target $Re_\tau$ \\
\midrule
I & 5600 & 5050 & 160 \\
II & 5050 & 5600 & 180 \\
III & 5600 & 9000 & 270 \\
IV & 5600 & 13752 & 395 \\
V & 5600 & 21944 & 590 \\
\bottomrule
\end{tabular}
\caption{Setup of the oRANS channel flow numerical experiments. Each case is defined by the bulk Reynolds number $Re_b$ of the periodic reference DNS from which the inflow fluctuations are rescaled (second column) (the same $Re_\tau=180$ simulation for cases I and III--V, and the $Re_\tau=160$ simulation for case II), the target bulk Reynolds number and the corresponding target friction Reynolds number $Re_\tau$. The experiments span a range of $Re_\tau$ from 160 to 590 to test model performance across increasing Reynolds numbers.}
\label{tab:onlineparams}
\end{table}

\begin{table}
\centering
\begin{tabular}{lccccccc}
\toprule
& & & \multicolumn{5}{c}{oRANS, $L_x=$} \\
\cmidrule(lr){4-8}
Case ($Re_\tau$) & $k$--$\omega$ & Offline DL & $4\pi\delta$ & $2\pi\delta$ & $\tfrac{4}{3}\pi\delta$ & $\tfrac{2}{3}\pi\delta$ & $\tfrac{1}{2}\pi\delta$ \\
\midrule
I (160)   & 1.0 & 0.12 & 0.13 & 0.10 & 0.09 & 0.09 & 0.09 \\
II (180)  & 1.0 & 0.23 & 0.16 & 0.26 & 0.21 & 0.28 & 0.30 \\
III (270) & 1.0 & 0.24 & 0.20 & 0.21 & 0.19 & 0.48 & --   \\
IV (395)  & 1.0 & 0.49 & 0.16 & 0.13 & 0.55 & 0.16 & --   \\
V (590)   & 1.0 & 1.17 & 0.14 & 0.13 & 0.89 & --   & --   \\
\bottomrule
\end{tabular}
\caption{Normalised cost functional $J^*$ of velocity and turbulent kinetic energy for baseline RANS ($k$--$\omega$), offline optimised RANS (trained on full-domain DNS statistics at $Re_\tau=180$), and online optimised RANS (oRANS) with different streamwise domain fractions $L_x$. Values below $1$ indicate improvement over the baseline RANS. Entries marked “--” correspond to cases where oRANS training diverged due to insufficient inflow length.}
\label{tab:wmse_results}
\end{table}

\subsubsection{Results summary}

Table~\ref{tab:wmse_results} summarises the training results across computed Reynolds numbers and channel lengths, expressed in terms of the normalised cost functional (equation \ref{eq:jstar}). The results show that the baseline $k$--$\omega$ RANS model exhibits large errors, particularly at higher $Re_\tau$. In all cases, this is largely due to a misprediction in the location and magnitude of peak turbulent kinetic energy in the channel. Offline DL training on the full periodic domain significantly improves predictions compared to baseline RANS, but the generalisation to out-of-sample Reynolds numbers is limited. In contrast, the online-optimised RANS approach achieves generally lower errors across all Reynolds numbers, even when trained with relatively small embedded subdomains. Accurate training requires only modest inflow lengths ($L_x \approx 2\pi\delta$), but sensitivity to domain size becomes more pronounced at higher $Re_\tau$. For very short subdomains ($L_x \leq \tfrac{2}{3}\pi\delta$), oRANS performance significantly degrades, and in some far out-of-sample cases training diverges entirely (indicated by ``--'' in the table), primarily due to boundary condition contamination.

\subsubsection{Velocity and turbulent kinetic energy profiles}

\begin{figure}
    \centering
    \includegraphics[width=\textwidth]{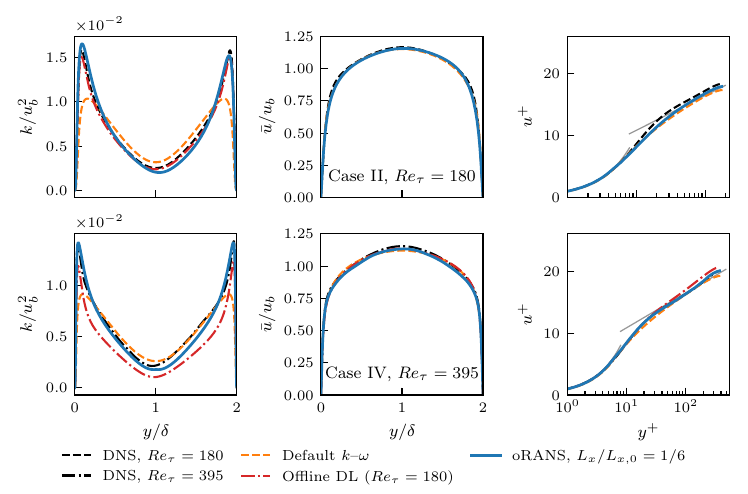}
    \caption{Turbulent kinetic energy (left), mean velocity (centre) and inner-scaled mean velocity (right) for cases II ($Re_\tau=180$, top) and IV ($Re_\tau=395$, bottom) at $L_x/L_{x,0}=1/6$: default $k$--$\omega$, offline-trained DL--RANS (trained on the $Re_\tau=180$ DNS), oRANS, and the reference DNS of each case; $u_b$ is defined in section~\ref{sec:ransnumerics}. In the right column each profile is scaled with its own friction velocity, so that an error in wall shear stress displaces the profile; grey lines are $u^+=y^+$ and the logarithmic law.}
    \label{fig:Online}
\end{figure}

To better illustrate the error mechanisms underlying these summary metrics, we present example velocity and turbulent kinetic energy profiles of cases II, IV for $L_x/L_{x,0} = \frac{1}{6}$ in figure \ref{fig:Online} comparing the differences in RANS, traditional offline supervised physics informed ML workflows and oRANS. All other cases show comparable profile distributions. 

For the channel flow cases considered, all three models reproduce the outer-scaled mean velocity profile with good accuracy, and the differences appear in $k$ and in inner scaling (left and right columns of figure~\ref{fig:Online}). The default model underpredicts the peak of $k$ by 35\% at both Reynolds numbers. oRANS recovers it to within 4\% at $Re_\tau=180$ and 1\% at $Re_\tau=395$, whereas the offline model, trained at $Re_\tau=180$, recovers it at $Re_\tau=180$ and underpredicts it by 13\% at $Re_\tau=395$. In the right column each profile is scaled with its own friction velocity, so that an error in wall shear stress displaces the profile. oRANS recovers the wall shear stress of the DNS to within 2\% at both Reynolds numbers, improving on the default model at $Re_\tau=180$, where it overpredicts the stress by 4.5\%, and matching it at $Re_\tau=395$. The offline model underpredicts the stress by 4\% at $Re_\tau=395$ and its $u^+$ profile is displaced upward through the logarithmic region. Agreement of the outer-scaled mean profile is in any case not expected to generalise to more complex flows, where RANS closures are known to mispredict the mean profile \citep{Wu2019}. The turbulent kinetic energy distribution is however poorly predicted by the default RANS model, with significant underproduction in the buffer to log-layer transition and mild overproduction in outer layer. For the mildly out-of-sample case II, offline supervised RANS-ML accurately corrects this error, and oRANS performance is comparable. In the far out-of-sample case IV however, the offline supervised workflow begins to fail but still outperforms the RANS. oRANS comparatively successfully recovers the distribution, particularly in the peak production region where  offline supervised RANS-ML begins to deviate from the DNS.

More broadly, out-of-sample degradation in offline training is expected to be even more pronounced for turbulence models that rely more heavily on machine learning; for instance, when the entire Reynolds stress tensor is represented as a neural network output. In such cases, the lack of physical anchoring increases sensitivity to distributional shift, and the importance of embedded physics-informed strategies is expected to be more pronounced.

\subsubsection{Effect of subdomain length}\label{sec:subdomain_length}

\begin{figure}
    \centering
    \includegraphics[width=\textwidth]{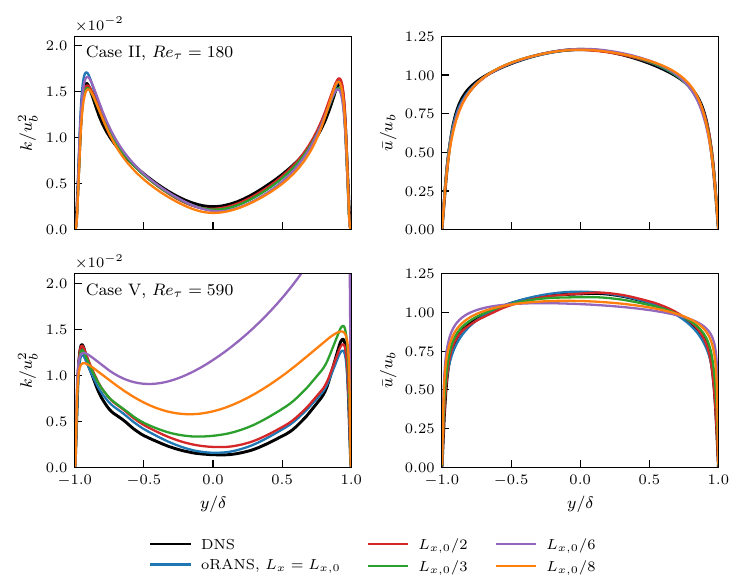}
    \caption{Comparison of turbulent kinetic energy $k$ (left) and mean velocity $\bar{u}$ (right) profiles for cases II ($Re_\tau=180$) and V ($Re_\tau=590$) with varying streamwise subdomain lengths $L_x$ used for oRANS training. At $Re_\tau=180$, oRANS predictions remain accurate across all subdomain lengths. At $Re_\tau=590$, agreement with DNS degrades as $L_x$ is reduced, with excessively short domains ($L_x/\delta = \tfrac{2}{3}\pi$ and $\tfrac{1}{2}\pi$) exhibiting algorithmic divergence due to boundary condition data contamination.}
    \label{fig:LL0}
\end{figure}

We now turn from comparing models to examining the role of domain size in oRANS performance. Consider the effect of shorter streamwise domains on oRANS, shown in figure \ref{fig:LL0} for weakly out-of-sample ($Re_\tau=180$) and far out-of-sample ($Re_\tau=590$) cases.

For the weakly out-of-sample case, oRANS performs well even with very short domains, although the turbulent kinetic energy is mildly underpredicted in the outer layer. In contrast, for the far out-of-sample case, the limitations of short domains become evident: training diverges as domain length decreases. This deterioration is also seen in the mean velocity profiles, which fall below even baseline RANS predictions. The failure arises due to a self-reinforcing instability associated with boundary-condition contamination. For far out-of-sample cases, both the inlet and outlet boundaries are less likely to closely respect the Navier--Stokes dynamics. Because the governing equations are elliptic in nature, boundary errors propagate throughout the domain, with the strongest impact in the vicinity of the boundaries. These polluted statistics are then recycled into the training, amplifying the error and ultimately driving divergence. These results indicate that while modest embedded domains suffice for training, excessively short domains cannot be used reliably for out-of-sample predictions. More generally, the appropriate domain size is problem-specific, and accurate boundary-condition representation is crucial for robust oRANS performance.

\begin{figure}
    \centering
    \includegraphics[width=\textwidth]{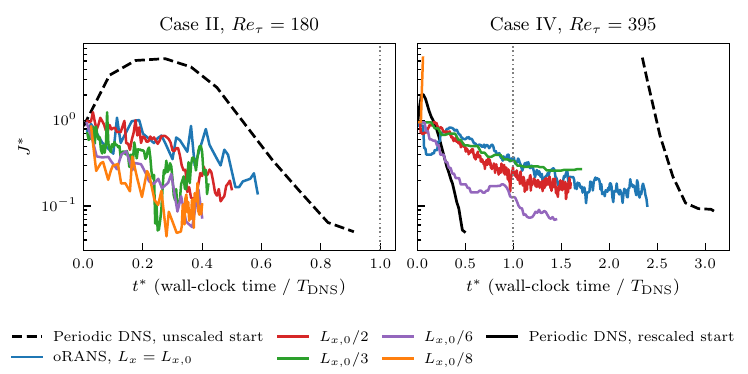}
    \caption{Convergence of the normalised cost functional $J^*$ with $t^*$, the wall-clock time scaled by that of the transient-plus-averaging window $T_{\mathrm{DNS}}$ of the reference periodic DNS, for cases II ($Re_\tau=180$, left) and IV ($Re_\tau=395$, right). Black: the error of the running statistics of the periodic DNS itself, started from an unscaled, perturbed field (left, and right dashed) or from the $Re_\tau=180$ field with the fluctuations rescaled by equation~\ref{eq:rescale} (right, solid). Colours: oRANS with $L_x/L_{x,0}=1$ to $1/8$.}
    \label{fig:convergence}
\end{figure}

\begin{table}
\centering
{\small
\begin{tabular}{lcccccc}
\toprule
& periodic DNS & \multicolumn{5}{c}{oRANS, $L_x/L_{x,0}=$} \\
& & $1$ & $1/2$ & $1/3$ & $1/6$ & $1/8$ \\
\midrule
$Re_\tau=180$, DNS unscaled start ($t^*_{\mathrm{DNS}}=0.72$) & 1 & 0.71 & 0.48 & 0.33 & 0.31 & 0.18 \\
$Re_\tau=395$, DNS rescaled start ($t^*_{\mathrm{DNS}}=0.37$) & 1 & 3.57 & 2.62 & 3.88 & 1.24 & -- \\
\phantom{$Re_\tau=395$,} DNS unscaled start ($t^*_{\mathrm{DNS}}=2.80$) & 1 & 0.47 & 0.34 & 0.51 & 0.16 & -- \\
\bottomrule
\end{tabular}}
\caption{Wall-clock time to reach $J^*=0.2$, a fivefold reduction of the baseline error, relative to that of the reference periodic DNS, for cases II and IV. At $Re_\tau=180$ the DNS is started from an unscaled, perturbed field; two DNS references are given at $Re_\tau=395$: started from the $Re_\tau=180$ field with the fluctuations rescaled by equation~\ref{eq:rescale}, the transformation used for the oRANS inflow, and started from the same field unscaled, which is the ordinary practice. `--': the run diverged (table~\ref{tab:wmse_results}).}
\label{tab:channel_cost}
\end{table}

\begin{figure}
    \centering
    \includegraphics[width=\textwidth]{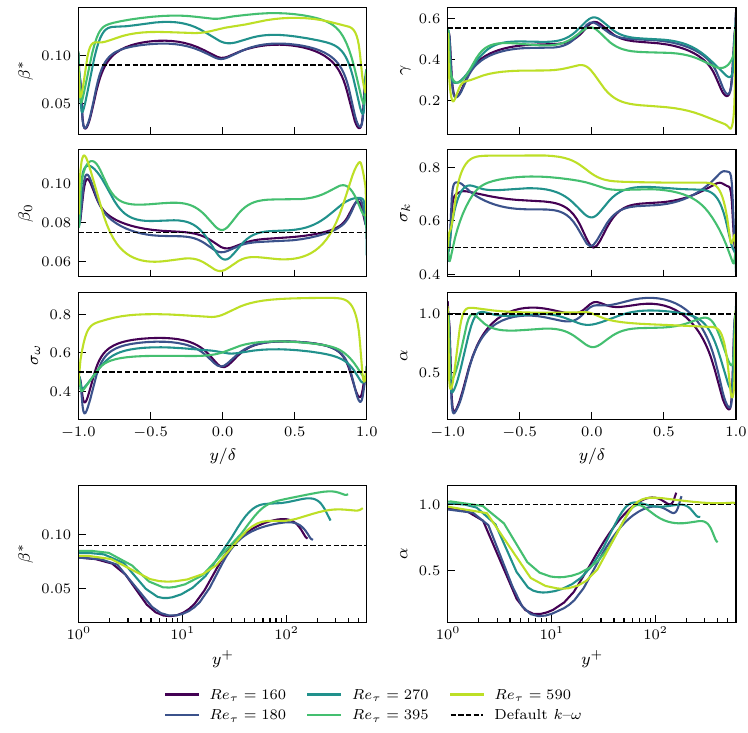}
    \caption{Learned closure functions of the online-trained models with $L_x/L_{x,0}=1/6$ (the runs of figure~\ref{fig:Online}): wall-normal profiles of $\beta^*$, $\gamma$, $\beta_0$, $\sigma_k$, $\sigma_\omega$ and $\alpha$ (with $\nu_t=\alpha k/\omega$) for $Re_\tau=160$ to $590$, dark to light. Bottom row: $\beta^*$ and $\alpha$ in the lower half of the channel against $y^+$. Dashed lines: the constants of the standard $k$--$\omega$ model. The variation of the learned terms illustrates how the closure adapts to different flow regimes across the channel.}\label{fig:ransparameters}
\end{figure}

\subsubsection{Convergence and computation time}\label{sec:convergence_cost}
We present the convergence history of oRANS and periodic DNS for weakly and significantly out-of-sample cases in figure \ref{fig:convergence}. Here $t^*$ denotes the normalised simulation time, defined by scaling the wall-clock time by the characteristic convergence window of the reference DNS. At $Re_\tau=180$, the baseline DNS exhibits a prolonged transient in which the weighted error initially grows before eventual decay, whereas oRANS rapidly settles: the rescaled fluctuations yield near-monotonic convergence across the embedded domain lengths. A periodic DNS started from a perturbed field must develop the whole flow, and its slowest parts are the mean profile and the energy-containing eddies, which develop on the outer timescale. In oRANS the inflow supplies both from the RANS solution, and the embedded region has then to produce only the turbulence consistent with them. Each scale $\ell$ adjusts in about its own turnover time $\tau(\ell)=\ell/u(\ell)$, where $u(\ell)$ is the velocity of the motions of that scale. For a spectrum $E(k)\sim k^{-n}$ the energy at scales up to $\ell$ is $u(\ell)^2\sim\ell^{\,n-1}$, so $\tau(\ell)\sim\ell^{(3-n)/2}$ increases with $\ell$ for any spectrum shallower than $k^{-3}$, and in the inertial range as $\ell^{2/3}\varepsilon^{-1/3}$ \citep[\S6.1]{pope2000}. The small and moderate scales therefore equilibrate within a short distance of the inlet (section~\ref{sec:rep_channel}, figure~\ref{fig:stats_8.72_df6}). For the same reason a closure can be trained from a region shorter than the largest scales of the flow. It is trained on the mean velocity and the turbulent kinetic energy. The largest scales enter the embedded region with the inflow, at the energy the RANS solution prescribes, and leave through the convective outflow, so their energy is present in the sampled statistics although the region is too short to regenerate them, and what the region regenerates is the small and moderate scales. In a periodic domain of the same length such structures cannot exist, and their production is forced into the scales that fit (section~\ref{sec:periodiceDNS}). For the more challenging $Re_\tau=395$ case, the same qualitative behaviour is observed, with oRANS showing near monotonic convergence, although it converges more slowly than at $Re_\tau=180$ owing to the longer streamwise development and feedback delay at higher $Re_\tau$ together with a larger mismatch between the inlet rescaling and the true statistics. As shown in figure \ref{fig:convergence}, for $Re_\tau=395$ the periodic DNS converges substantially slower if restarted directly from the baseline DNS field without rescaling. The computational burden is dominated by the embedded simulation, as the RANS model and its adjoint are solved only once per parameter update, every $M$ time steps of the embedded simulation (section~\ref{sec:implementation}), and can be evaluated in parallel. Consequently, the per-timestep cost of oRANS scales approximately linearly with the embedded domain fraction, $L_x/L_{x,0}$. Shorter domains further reduce the effective feedback time, since the statistics are sampled closer to the inlet and the characteristic flowthrough time is smaller. 

However, if the statistics are sampled too close to the inlet, the statistics are contaminated by transient adjustment effects, which introduce noise into the training and can misdirect the gradient. This tradeoff highlights both the efficiency and the limitations of short-domain training in oRANS.
Table~\ref{tab:channel_cost} compares the cost of the online closure with that of the reference DNS, both measured as wall-clock time to reach $J^*=0.2$, a fivefold reduction of the baseline error. The comparison is made at fixed accuracy because the running statistics of the periodic DNS converge to the reference itself. Against a periodic DNS started in the ordinary way, from an unscaled field, the online closure reaches this accuracy in $18$--$71\%$ of the DNS time at $Re_\tau=180$ and $16$--$47\%$ at $Re_\tau=395$ for the embedded regions that reach it. The $L_{x,0}/8$ region diverges at $Re_\tau=395$. Against a periodic DNS started from the $Re_\tau=180$ field with the fluctuations rescaled by equation~\ref{eq:rescale}, the transformation the oRANS inflow itself uses, no embedded region is faster at $Re_\tau=395$. That start shortens the transient of the periodic DNS nearly eightfold, whereas the online training must first adapt the closure from the $Re_\tau=180$ reference, and figure~\ref{fig:convergence} shows the DNS converging within a few tenths of a window while the errors of the closures are still decreasing.

The method is preferable to LES when the computational domain is much larger than the region that must be resolved to capture the dominant turbulence. In practice the embedded region would itself be resolved by LES or wall-modelled LES, so the method confines that simulation to a part of the domain and uses RANS with the trained closure elsewhere, and the saving relative to such a simulation of the whole domain is the fraction of the domain left unresolved. In the channel the embedded region is a sizeable fraction of a domain only a few integral scales long, and the channel serves to test the method against a converged reference. The saving is large in domains spanning many integral scales of which the embedded region covers one or a few, such as the wind farm discussed in section~\ref{sec:rep_channel}, and in flows without reference data, where the alternative to online training is an uncalibrated closure.

Once trained, the closure is an ordinary $k$--$\omega$ model with a network for its coefficients and is evaluated at the cost of the RANS solve alone. This distinguishes the method from hybrid RANS--LES, in which the resolved region is placed where accuracy is required and supplies the solution there for the whole calculation. Here the resolved region calibrates a closure that is applied across the RANS domain and is required only until the closure has converged, so the method suits flows in which the turbulence of the embedded region is representative of that elsewhere.

Finally, we note that oRANS introduces additional I/O overhead compared to periodic DNS. In the present implementation, the reference DNS dataset must be retained in memory, and the replay buffer must be saved on write. The buffer also introduces communication overhead, which may become significant at scale, though it is minor compared to the DNS cost in the present setup.

\subsection{Turbulence representation}\label{sec:rep_channel}

To better understand how oRANS adapts to different flow regimes, we briefly analyse detailed turbulent statistics for example training flows.

First consider the learned turbulence model coefficients across Reynolds numbers. Figure~\ref{fig:ransparameters} shows the wall-normal variation of the key closure terms $\alpha$, $\beta_0$, $\beta^*$, $\gamma$, $\sigma_k$, and $\sigma_\omega$ obtained from the online-trained models. The dashed lines denote the constant baseline values used in the standard $k$--$\omega$ model. Three features are common to the converged runs. First, in the buffer layer both the dissipation coefficient of $k$ and the eddy-viscosity coefficient are reduced well below their standard values, with minima near $y^+=10$ at every Reynolds number (bottom row of figure~\ref{fig:ransparameters}), and both recover their outer values by the start of the logarithmic layer. The learned near-wall correction therefore scales in wall units. This corresponds to the viscous damping that the low-Reynolds-number forms of the $k$--$\omega$ model impose through functions of the turbulence Reynolds number $k/\nu\omega$ \citep{Wilcox1998}, one of the inputs of the closure, and is recovered here from the sampled statistics. Its effect is the larger near-wall peak of $k$, without a corresponding rise of $\nu_t$, that the default model lacks (figure~\ref{fig:Online}). Second, across the logarithmic and outer regions the production coefficient of $\omega$ is lowered to $\gamma\simeq0.4$ and the diffusion coefficients raised to $\sigma_k\simeq0.7$ and $\sigma_\omega\simeq0.6$, while $\beta^*$ rises above its standard value in the core. In a logarithmic layer the coefficients of the $k$--$\omega$ model are constrained by $\gamma=\alpha\beta_0/\beta^*-\sigma_\omega\kappa^2\sqrt{\alpha/\beta^*}$, the relation that fixes $\gamma=5/9$ for the standard constants with $\kappa=0.41$, and the learned outer values imply $\kappa\simeq0.44$ for $Re_\tau=160$ to $395$. The coefficients therefore change together in a way that retains a logarithmic law. The change is towards the outer-layer values that blended models adopt \citep{Menter1994}, but is obtained without a blending function. Third, the near-wall corrections collapse in $y^+$ and the outer corrections in $y/\delta$, so the learned coefficients reflect the two-layer structure of wall turbulence.

A key distinction emerges between offline-trained models and oRANS. An offline-trained closure is fixed once trained and is not recalibrated at the target Reynolds number. In contrast, the online optimisation modifies the coefficients in a Reynolds-number-dependent way, reflecting changes in turbulence production and transport. These adjustments are not imposed \textit{a priori} but arise naturally from the coupled training with embedded statistics, suggesting that the learned corrections adapt to changes in large-scale turbulence dynamics.

These variations show that the learned coefficients acquire non-trivial wall-normal and Reynolds-number dependence. By breaking the rigidity of constant-coefficient closures, oRANS reshapes the turbulence representation in response to the target flow, which a closure trained offline, before the target flow is computed, does not do.

\begin{figure}
    \centering

\begin{subfigure}[t]{0.45\textwidth}
  \centering
  \begin{overpic}[width=\linewidth]{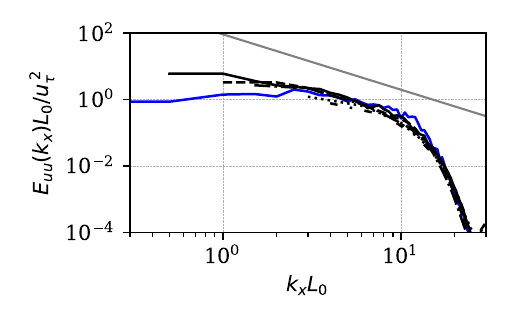}
    \put(30,24){\fcolorbox{black}{white}{\small $Re_{\tau}=180$}}
  \end{overpic}
\end{subfigure}\hfill
\begin{subfigure}[t]{0.45\textwidth}
  \centering
  \begin{overpic}[width=\linewidth]{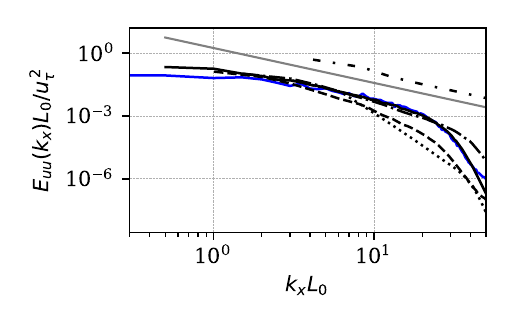}
    \put(30,24){\fcolorbox{black}{white}{\small $Re_{\tau}=395$}}
  \end{overpic}
\end{subfigure}
    \begin{subfigure}[t]{0.45\textwidth}
        \centering
        \includegraphics[width=\linewidth]{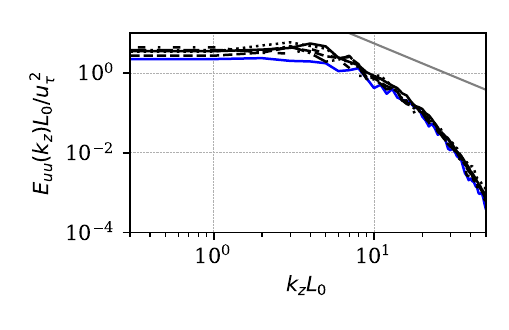}
    \end{subfigure}%
    \hfill
    \begin{subfigure}[t]{0.45\textwidth}
        \centering
        \includegraphics[width=\linewidth]{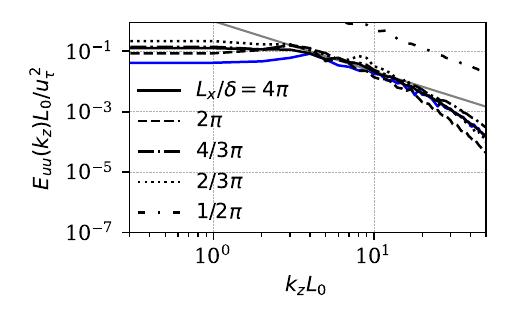}
    \end{subfigure}

    \begin{subfigure}[t]{0.45\textwidth}
        \centering
        \includegraphics[width=\linewidth]{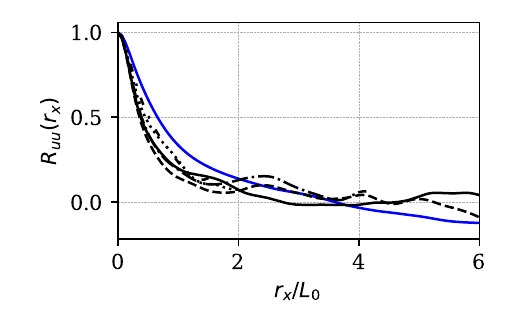}
    \end{subfigure}%
    \hfill
    \begin{subfigure}[t]{0.45\textwidth}
        \centering
        \includegraphics[width=\linewidth]{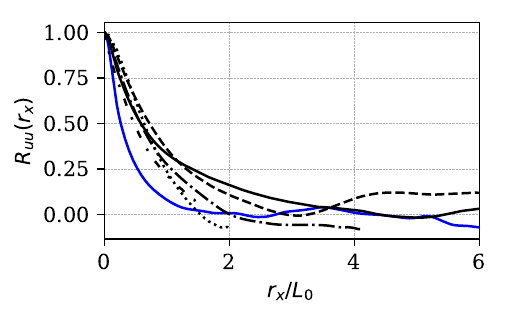}
    \end{subfigure}

    \caption{
Turbulence statistics for oRANS case II ($Re_\tau=180$, left) and case IV ($Re_\tau=395$, right) at $y/\delta=0.829$. Top row: streamwise spectra $E_{uu}(k_x)$. Middle row: spanwise spectra $E_{uu}(k_z)$. Bottom row: longitudinal two-point correlations $R_{uu}(r_x)$. Blue curves show reference periodic DNS data, and black curves correspond to the embedded simulation with domain lengths $L_x=4\pi\delta$ (solid), $2\pi\delta$ (dashed), $\tfrac{4}{3}\pi\delta$ (dash-dot), $\tfrac{2}{3}\pi\delta$ (dotted), and $\tfrac{1}{2}\pi\delta$ (loosely dashed). As $L_x$ decreases, $k_{\min}=2\pi/L_x$ increases and wavelengths longer than $L_x$ are represented only in part, producing the observed loss of large-scale content and shorter correlation lengths. Divergence is seen in the shortest domains, where boundary-condition contamination dominates.}
    \label{fig:spectra_outer_comparison}
\end{figure}

\begin{figure}
    \centering
    \begin{subfigure}{0.95\textwidth}
        \centering
        \includegraphics[width=\textwidth]{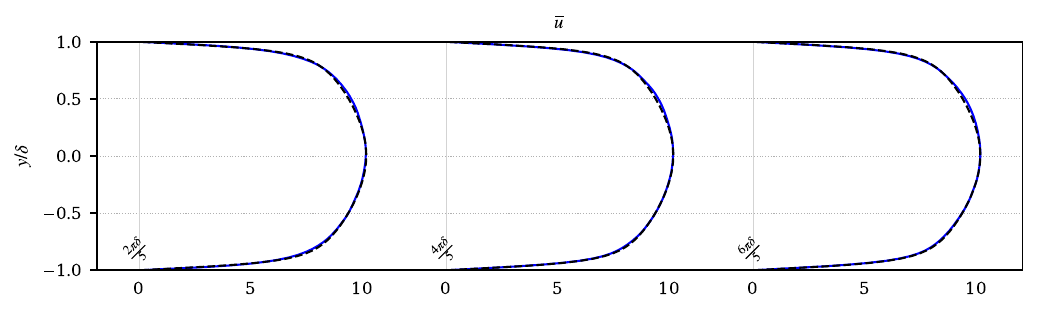}
    \end{subfigure}
    
    \begin{subfigure}{0.95\textwidth}
        \centering
        \includegraphics[width=\textwidth]{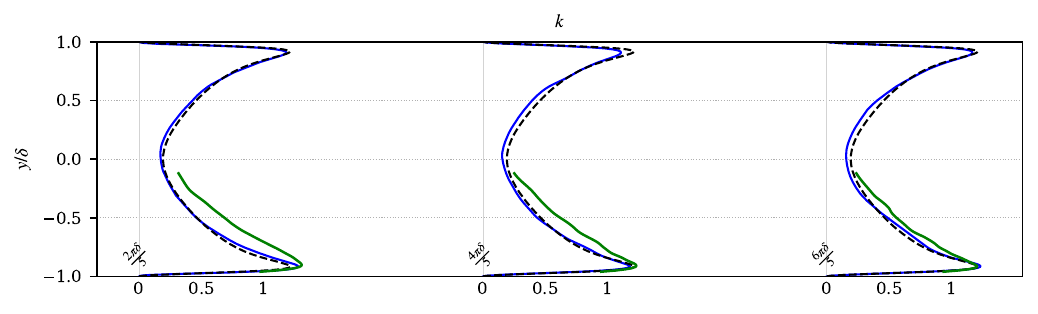}
    \end{subfigure}
    
    \begin{subfigure}{0.95\textwidth}
        \centering
        \includegraphics[width=\textwidth]{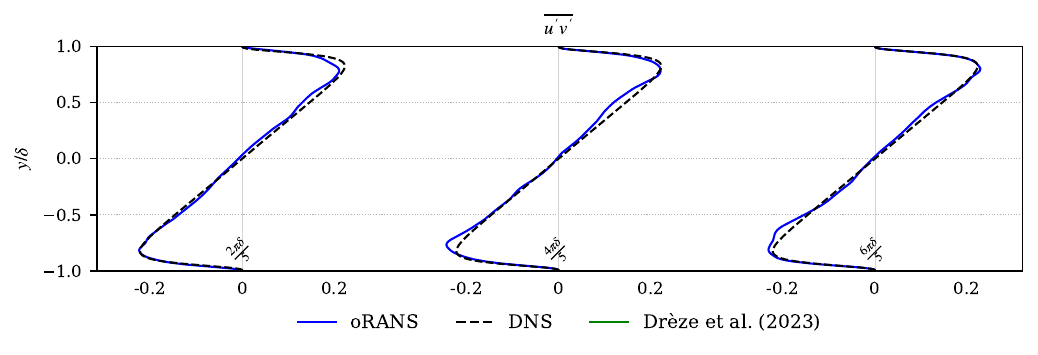}
    \end{subfigure}
    
    \caption{
    Streamwise evolution of mean flow statistics for case II ($Re_\tau=180$), and for a full length embedded domain $L_x = 4\pi\delta$. From top to bottom: (a) mean streamwise velocity $\bar{u}$, (b) turbulent kinetic energy $k$, and (c) Reynolds shear stress $\overline{u'v'}$, each as a function of the streamwise coordinate $x$. Results from oRANS (blue), DNS (black dashed), and \citet{Dreze2023} (green) are shown. The oRANS predictions recover the true DNS statistics earlier in the streamwise direction compared to \cite{Dreze2023}, highlighting the effectiveness of the rescaled turbulent inflow procedure.
    }
    \label{fig:stats_8.72_df6}
\end{figure}

Next, figure \ref{fig:spectra_outer_comparison} compares the streamwise spectra and two-point velocity correlations against the fully periodic case, for example, case II.
The channel flow spectra highlight an important limitation of the embedded-simulation strategy: whenever the high-fidelity sub-domain is shorter than the largest energetic structures, the low-wavenumber content of the spectrum is represented only in part, and the optimisation of the $k$-equation inevitably degrades. In a truncated embedded region the very low streamwise wavenumbers, which originate in the channel centre where the largest eddies reside, are represented only in part, and this is precisely where the relative error in the optimised $k$ profile becomes visible. The tests with short embedded regions therefore constitute an especially stringent, ``worst-case'' scenario for oRANS; that the method still yields reasonable agreement is encouraging.
\par
At the same time, the high-wavenumber range of the spectra, the cascade, and the two-point correlations remain accurately reproduced even for modest embedded-region lengths. This indicates that oRANS faithfully captures small-scale turbulent dynamics, while its limitations are confined to the very largest structures excluded by the truncated domain.

Finally, we examine the streamwise evolution of turbulence statistics in short embedded domains. Figure~\ref{fig:stats_8.72_df6} shows the example case at $Re_\tau=180$.
Across the Reynolds-numbers tested, we find that the re-scaled fluctuations recover statistically stationary turbulence within one integral length downstream of the inlet: in cases where oRANS matches DNS (see table \ref{tab:wmse_results}) the mean profiles of $u$, $k$, and $\overline{u'v'}$ are already indistinguishable from fully developed DNS data by $x = 1.25\delta$. This is markedly faster than state-of-the-art synthetic-inflow techniques such as the RANS-guided method of \citet{Dreze2023}, which reach comparable accuracy only after $x\approx6\delta$. The key difference is that oRANS feeds the embedded sub-domain with physically consistent, dynamically evolving fluctuations from an existing DNS dataset rather than statistically prescribed surrogates, which require a longer relaxation distance but have the advantage of only requiring a baseline RANS solution. The limitation of the approach is equally clear: if the embedded region is shorter than the lowest streamwise wavenumbers, the largest motions are only partly contained within it and are inherited from the inflow rather than regenerated. In this case, higher-order moments converge quickly, but their equilibrium differs from the true mean depending on the domain length. This represents a hard constraint on accuracy that cannot be overcome by training alone.

\par
From an application perspective, this restriction may be mild: in many engineering configurations, the largest eddies are of the order of a characteristic geometry scale, while the computational domains span many such scales. For example, in a wind-farm setting, an LES patch of one or two rotor diameters embedded in a RANS domain covering dozens of turbines would resolve the relevant large-scale structures. In such cases the tests with longer embedded regions suggest that oRANS would recover the global statistics with high fidelity. Nevertheless, careful consideration of the largest turbulent scales relative to the chosen embedded patch remains essential when extending oRANS beyond canonical channel flow.

\section{Spanwise-rotating channel flow}\label{sec:rotating}

In the channel of section~\ref{sec:nsresults} the baseline closure contains the relevant physics, and the online optimisation refines its coefficients. We now consider fully developed channel flow rotating about the spanwise axis \citep{Johnston1972,Kristoffersen1993}, whose governing physics the baseline closure is structurally incapable of representing. Rotation enters the momentum equations of the embedded simulation through the Coriolis acceleration, but it vanishes identically from the $k$--$\omega$ equations, so a linear eddy-viscosity closure without a rotation correction returns the non-rotating solution at every rotation rate, regardless of its coefficients. The flow itself responds strongly, with turbulence augmented on the pressure side and suppressed on the suction side, and in the core the mean velocity takes the linear form $\partial_y\bar u\simeq2\Omega$ of vanishing absolute vorticity. Rotation corrections that restore part of this behaviour exist \citep{Spalart1997,PetterssonReif1999}, but each requires its own calibration, and \citet{Shur2000} remark that such corrections are ``more or less successful in the specific flows for which they were designed'' but ``are not universal''. The case is therefore a controlled test of the premise of oRANS, that the embedded simulation can supply physics the closure does not contain. Here the missing physics is known and can be checked against the DNS, whereas in the flows for which the method is intended it would not be known in advance. We examine whether an embedded simulation seeded with non-rotating turbulence, coupled to a closure with no rotation term and no rotation rate among its inputs, can recover the rotation response online, and what the learned closure then represents.

Starting from the non-rotating $k$--$\omega$ solution and non-rotating inflow fluctuations, oRANS reduces the error of the baseline closure by up to two thirds at $Ro_b=0.15$ (defined in section~\ref{sec:rot_setup}) and by up to 87\% at $Ro_b=0.5$, more than either established rotation correction, whereas a closure trained offline on the non-rotating DNS recovers none of the response. The closure does so by learning the eddy-viscosity distribution of the rotating flow, raising it on the pressure side and reducing it to the order of the kinematic viscosity on the suction side, where the Reynolds shear stress nearly vanishes. The embedded region corrects the inflow, which carries the structure of non-rotating turbulence and the amplitude of the model being trained, towards the rotating state, and what it does not regenerate within the embedded region is inherited by the closure. This is the boundary-condition contamination of section~\ref{sec:nsresults}, here with inflow turbulence taken from a different flow, and, as in the channel, it limits the accuracy of the shortest embedded regions.

\subsection{Governing equations with system rotation}\label{sec:rot_equations}

System rotation enters the embedded simulation only through the forcing term of equation~\eqref{eq:momentum}. In a frame rotating at angular velocity $\boldsymbol\Omega=(0,0,\Omega)$,
\begin{equation}\label{eq:rot_momentum}
f_i = F\,\delta_{i1} - 2\,\epsilon_{ijk}\Omega_j u_k,
\end{equation}
the driving pressure gradient $F$ together with the Coriolis acceleration $(2\Omega v,\,-2\Omega u,\,0)$, with the centrifugal acceleration absorbed into the pressure. Averaged over $x$, $z$ and time, the Coriolis term vanishes from the mean streamwise momentum because $\bar v=0$, and is balanced by the mean pressure in the wall-normal direction, so that, in dimensional variables,
\begin{equation}\label{eq:rot_mean}
0 = F + \frac{\mathrm d}{\mathrm d y}\left[\nu\frac{\mathrm d\bar u}{\mathrm d y}
- \overline{u'v'}\right]
\end{equation}
is the mean momentum equation of the non-rotating channel. Rotation acts on the turbulence instead, through a production term in the Reynolds-stress transport equations that is trace-free \citep{Johnston1972}. It therefore appears neither in the equation for $k=\tfrac12\overline{u_i'u_i'}$ nor in the $\omega$ equation modelled from it. The RANS equations~\eqref{eq:RANS} with a linear eddy-viscosity closure are unchanged by rotation. Any closure whose coefficients are functions of the local invariants~\eqref{eq:local} of the RANS state returns the same solution at every $Ro_b$, the non-rotating one for a closure calibrated without rotation. Rotation can enter such a closure only through the data it is trained against.

The effect of rotation on the turbulence is governed by the ratio of system to mean-shear vorticity,
\begin{equation}\label{eq:rot_S}
S = -\frac{2\Omega}{\partial_y\bar u},
\end{equation}
with $-1<S<0$ destabilising, $S>0$ stabilising and $S=-1$ the neutral state of zero absolute vorticity \citep{Bradshaw1969,Johnston1972}. For $\Omega>0$ and the present orientation, the wall at $y=-\delta$ is the pressure side, where the Coriolis force feeds the wall-normal fluctuations and the shear stress, and the wall at $y=+\delta$ is the suction side, where it drains them and the turbulence is suppressed. The classical corrections make the eddy-viscosity coefficient \citep{PetterssonReif1999} or the production term \citep{Spalart1997} a function of $S$ or of an equivalent invariant.

\subsection{Numerical experiments}\label{sec:rot_setup}

Following \citet{Kristoffersen1993}, the rotation rate is characterised by the rotation number $Ro_b = 2\Omega\delta/u_b$, and two values are considered. At $Ro_b=0.15$ both sides of the channel remain turbulent and the asymmetry is moderate. At $Ro_b=0.5$ the suction side is close to relaminarisation while the pressure side remains fully turbulent, and the departure from the non-rotating solution is largest. At higher rotation rates the pressure-side turbulence is damped in turn and the profiles approach a symmetric laminar form \citep{Grundestam2008}. For a rotation-blind closure, $Ro_b=0.5$ is therefore the hardest case, since its target lies furthest from the only solution the baseline can produce.

\begin{figure}
\centering
\includegraphics[width=\textwidth]{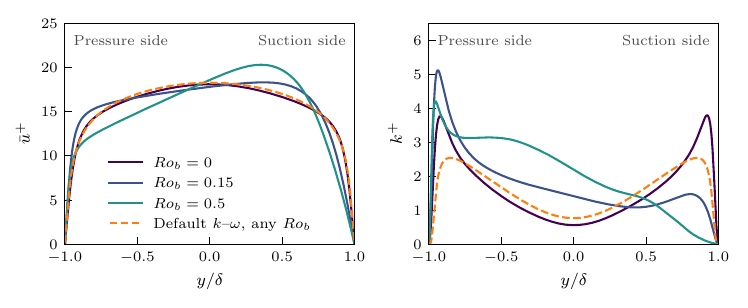}
\caption{Reference DNS of spanwise-rotating channel flow at $Re_\tau=194$
for $Ro_b=0$, $0.15$ and $0.5$: mean velocity (left) and turbulent kinetic
energy (right) in wall units, with the default $k$--$\omega$ solution,
which at a given bulk velocity is independent of $Ro_b$. The pressure side is at $y/\delta=-1$ and
the suction side at $y/\delta=1$.}
\label{fig:rot_setup}
\end{figure}

The reference DNS are driven by a constant mean pressure gradient at $Re_\tau=u_\tau\delta/\nu=194$, where $u_\tau=[(u_{\tau,p}^2+u_{\tau,s}^2)/2]^{1/2}$ is the friction velocity averaged over the two walls. The DNS are validated against \citet{Kristoffersen1993} in \ref{app:rot_validation}. Figure~\ref{fig:rot_setup} shows the mean velocity and turbulent kinetic energy beside the $k$--$\omega$ prediction, which at a given bulk velocity is the same at every $Ro_b$. As $Ro_b$ increases, the pressure-side peak of $k$ grows and the suction-side peak collapses, and the wall shear-stress ratio $\tau_{w,p}/\tau_{w,s}$ rises with $Ro_b$. The baseline predicts equal peaks and equal wall stresses at every rotation rate.

The oRANS configuration is that of section~\ref{sec:oranssetup}. The RANS model starts from the non-rotating $k$--$\omega$ solution and its pretrained network, and the inflow fluctuations are taken from the $Ro_b=0$ DNS and rescaled by equation~\eqref{eq:rescale} to the mean velocity and turbulent kinetic energy of the current RANS solution. The inflow therefore carries the amplitude profiles of the model being trained together with the structure of non-rotating turbulence, and at the outset it contains no information about rotation. As the closure learns, the rescaling passes its asymmetric mean and $k$ profiles to the inflow, but the structure of the fluctuations remains that of non-rotating turbulence. The rotating dynamics must develop within $\Omega_{\mathrm{eDNS}}$, and the closure can learn only what the region generates there. Training is continued until the online objective, smoothed over a few flow-throughs of the embedded region, reaches its minimum, and the state at that minimum is reported.

We compare against four references, the first of which is the baseline $k$--$\omega$ model. The second adds the rotation correction of \citet{Spalart1997}, in the form of \citet{Smirnov2009} with the original constants, which multiplies the production of $k$ and $\omega$ by a function of the local strain-to-vorticity ratio that exceeds unity on the pressure side and falls below it on the suction side. The third adds instead the correction of \citet{PetterssonReif1999}, which makes the eddy-viscosity coefficient $\alpha$ in $\nu_t=\alpha k/\omega$ an algebraic function of the strain and rotation invariants (their equation~26, with their constants), and reduces to $\alpha=1$ at $\Omega=0$. The fourth is an offline model with the same network, inputs and adjoint gradient as oRANS, trained on the non-rotating DNS and deployed at the rotating conditions, which uses only data that exist before the rotating flow has been computed.

\subsection{Results}\label{sec:rot_results}

\subsubsection{Results summary}

Starting from an inflow and a closure that contain no information about rotation, oRANS recovers most of the rotation response. Table~\ref{tab:rot_jstar} gives the normalised error at the end of training (section~\ref{sec:rot_setup}). With the full-length embedded region the error of the baseline closure falls by 56\% at $Ro_b=0.15$ and 81\% at $Ro_b=0.5$.

The offline model trained on the non-rotating DNS recovers none of the response. Deployed at the rotating conditions it is indistinguishable from the baseline at $Ro_b=0.15$ ($J^*=0.98$) and worse than it at $Ro_b=0.5$ ($J^*=1.20$). The only input in~\eqref{eq:local} that distinguishes the two sides of a channel is the sign of the shear $S^*$. A model trained on symmetric statistics learns an even function of $S^*$ and returns a symmetric solution at any rotation rate. The Spalart--Shur correction predicts the linear core and the displaced velocity maximum at $Ro_b=0.5$, with $k$ close to the baseline on both sides ($J^*=0.67$ and $0.85$). The Pettersson Reif correction predicts the displaced maximum and a lower suction-side $k$, with a suction-side peak the rotating flow does not have and no augmentation on the pressure side ($J^*=0.83$ and $0.55$). Every oRANS configuration in table~\ref{tab:rot_jstar}, including those sampled within the developing region, improves on the baseline, matches or improves on the Spalart--Shur model, and all but the shortest embedded region at $Ro_b=0.5$ on the Pettersson Reif correction.

\begin{table}
\centering
{\small\setlength{\tabcolsep}{4pt}
\begin{tabular}{lccccccccc}
\toprule
& $k$--$\omega$ & $k$--$\omega$ & $k$--$\omega$ & Offline DL & \multicolumn{5}{c}{oRANS, $L_x/L_{x,0}=$} \\
$Ro_b$ & & + S--S & + PR & $Ro_b{=}0$ & $1$ & $1/2$ & $1/3$ & $1/6$ & $1/8$ \\
\midrule
0.15 & 1.0 & 0.67 & 0.83 & 0.98 & 0.44 & 0.34 & 0.32 & 0.50 & 0.66 \\
0.5  & 1.0 & 0.85 & 0.55 & 1.20 & 0.19 & 0.16 & 0.13 & 0.50 & 0.72 \\
\bottomrule
\end{tabular}}
\caption{Normalised cost functional $J^*$ (equation~\ref{eq:jstar}) for the rotating channel. The third column is the $k$--$\omega$ model with the
Spalart--Shur correction (in the two-equation form of Smirnov \& Menter 2009, original constants) and the fourth with the eddy-viscosity correction of Pettersson Reif \emph{et al.} (1999). The offline model in the fifth column is trained on the
non-rotating DNS and deployed at the rotating conditions. All oRANS runs start from the non-rotating solution and inflow. S--S: the Spalart--Shur correction; PR: the correction of \citet{PetterssonReif1999}.}
\label{tab:rot_jstar}
\end{table}

\subsubsection{Velocity and turbulent kinetic energy profiles}

\begin{figure}
\centering
\includegraphics[width=\textwidth]{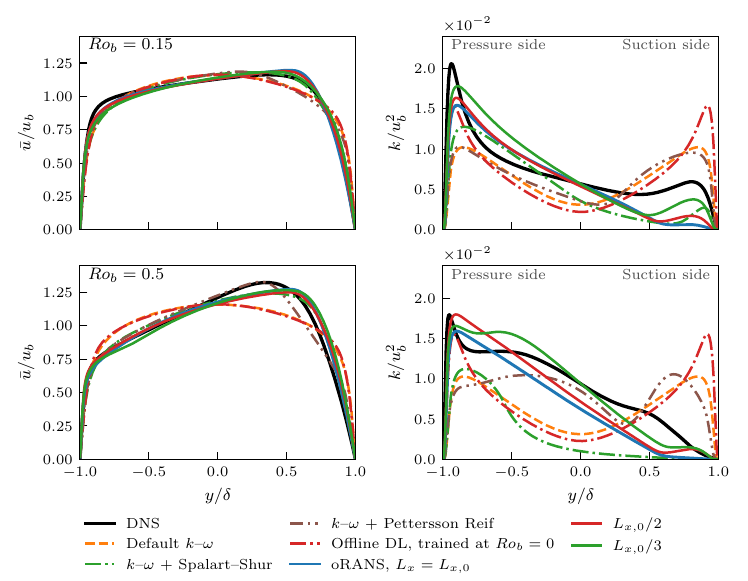}
\caption{Mean velocity (left) and turbulent kinetic energy (right) for $Ro_b=0.15$ (top) and $0.5$ (bottom): DNS, default $k$--$\omega$, $k$--$\omega$ with the Spalart--Shur correction \citep{Spalart1997} in the form of \citet{Smirnov2009} and with the Pettersson Reif correction \citep{PetterssonReif1999}, the offline model trained on the non-rotating DNS and deployed here (Offline DL), and oRANS at the end of training with $L_x=L_{x,0}$, $L_{x,0}/2$ and $L_{x,0}/3$, all started from the non-rotating solution and inflow. The $L_{x,0}/6$ and $L_{x,0}/8$ runs, whose statistics are sampled within the developing region (section~\ref{sec:rot_length}), are omitted for clarity. Their errors are in table~\ref{tab:rot_jstar}.}
\label{fig:rot_profiles}
\end{figure}

At both rotation numbers oRANS reproduces the asymmetry that the baseline cannot (figure~\ref{fig:rot_profiles}). Its $k$ peaks at the pressure wall and decreases across the channel, and its velocity maximum moves towards the suction side. The remaining error is mostly in $k$, whose pressure-side peak at $Ro_b=0.15$ is somewhat too low, and the weak secondary maximum that the DNS has near the suction wall is reproduced only partly. Section~\ref{sec:rot_length} shows that the embedded region does not regenerate this maximum from the non-rotating inflow turbulence.

At $Ro_b=0.5$ oRANS reproduces the linear core of slope $2\Omega$ of the DNS, its velocity maximum displaced towards the suction side, and the decay of its turbulent kinetic energy from the pressure-side peak to nearly zero on the suction side, where the flow is close to relaminarisation.

\begin{figure}
\centering
\includegraphics[width=\textwidth]{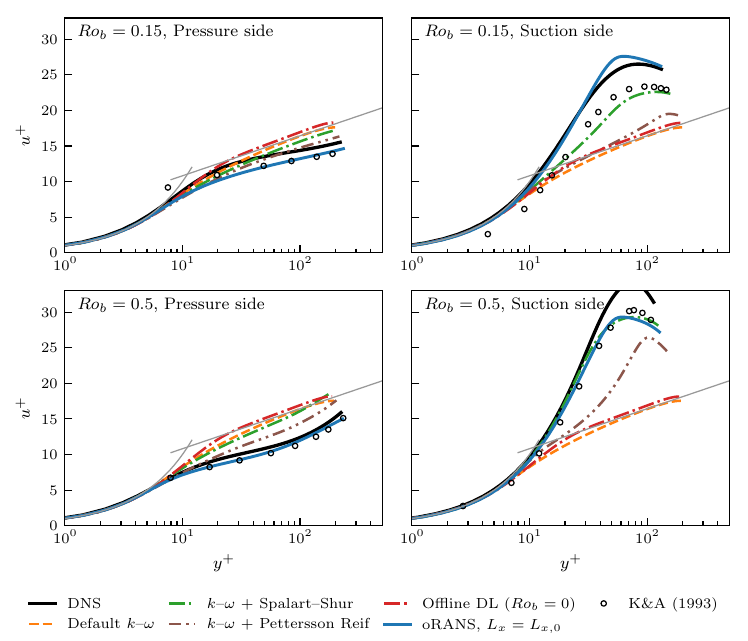}
\caption{Mean velocity in inner scaling with the local friction velocity
of each wall, pressure side (left) and suction side (right), for
$Ro_b=0.15$ (top) and $0.5$ (bottom): DNS, default $k$--$\omega$,
$k$--$\omega$ with the Spalart--Shur and Pettersson Reif corrections, the offline model
trained at $Ro_b=0$, oRANS with $L_x=L_{x,0}$ at its state at the end of training, and the
data of \citet{Kristoffersen1993}. Thin grey lines are $\bar u^+=y^+$ and
$\bar u^+=\ln y^+/0.41+5.2$.}
\label{fig:rot_uplus}
\end{figure}

Figure~\ref{fig:rot_uplus} shows the same velocity profiles in inner scaling, with the friction velocity of each wall. The rotating DNS departs from the logarithmic law on both sides. It lies below it on the pressure side, where the augmented mixing thickens the buffer layer, and far above it on the suction side, where the profile follows the viscous law well beyond the sublayer because the shear stress has collapsed. The rotation-blind closures return the logarithmic law on both sides at both rotation numbers. oRANS follows the DNS on both sides, closely at $Ro_b=0.5$ and with a small shortfall on the pressure side at $Ro_b=0.15$.

\subsubsection{Development along the embedded region and effect of its length}\label{sec:rot_length}

\begin{figure}
\centering
\includegraphics[width=\textwidth]{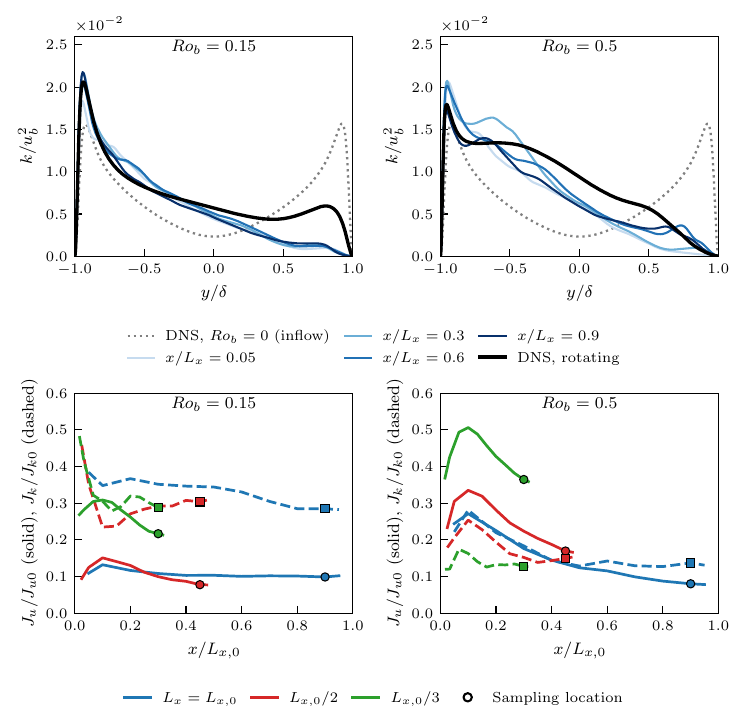}
\caption{Development of the embedded simulation along the embedded region at the end of training. Top: plane- and time-averaged $k$ at $x/L_x=0.05$, $0.3$, $0.6$ and $0.9$ (light to dark) for $L_x=L_{x,0}$, with the rotating DNS and the non-rotating DNS from which the inflow fluctuations are drawn. Bottom: error of the embedded statistics against the rotating DNS, $J_u/J_{u0}$ (solid) and $J_k/J_{k0}$ (dashed), at stations along the embedded region for $L_x=L_{x,0}$, $L_{x,0}/2$ and $L_{x,0}/3$, plotted against distance from the inlet in units of $L_{x,0}$. Symbols mark where the training statistics are sampled in each run. $Ro_b=0.15$ left, $0.5$ right.}
\label{fig:rot_planes}
\end{figure}

The embedded region converts the non-rotating inflow into the rotating state within the embedded region, and the closure inherits whatever the region does not regenerate before the statistics are sampled. As in the channel, the inflow supplies the mean and the energy, the region regenerates the small and moderate scales from them, and the scales longer than the region are represented only in part (figure~\ref{fig:spectra_outer_comparison}). The top row of figure~\ref{fig:rot_planes} gives, for $L_x=L_{x,0}$, the plane-averaged $k$ at stations along the embedded region, with the non-rotating DNS from which the inflow fluctuations are drawn and the rotating reference DNS. The bottom row gives, for the three longest regions, the error of the embedded statistics against the rotating DNS along the embedded region.

The mean-velocity error is small throughout, because the RANS mean imposed at the inlet is itself close to the rotating DNS once training has converged. For the full-length embedded region at $Ro_b=0.5$ it falls by a factor of three along the region, most of the decrease occurring within the first half of the embedded region.

At $Ro_b=0.5$ the inflow overshoots the pressure-side peak and carries too little energy on the suction side. Both errors relax towards the rotating DNS along the embedded region, and the suction-side $k$ nearly triples. At $Ro_b=0.15$ the pressure side and core are close to the DNS throughout, but the suction-side maximum of the rotating DNS, absent from the RANS $k$ imposed at the inlet, is barely regenerated within the region, so the closure is trained on statistics that under-represent it.

The same adjustment limits how short the embedded region can be. At $Ro_b=0.5$ the errors first rise over about $0.1L_{x,0}$ as the rescaled inflow adjusts to the embedded dynamics, and fall below their inlet values only after about $0.2L_{x,0}$. The $L_{x,0}/6$ and $L_{x,0}/8$ regions are sampled within this adjustment, so their statistics are those of a developing flow and the errors are reduced by only about $30$--$50\%$. At $Ro_b=0.15$ with $L_{x,0}/8$ the velocity error is not reduced, as in the far out-of-sample channel cases of section~\ref{sec:subdomain_length}.

\subsubsection{Convergence and computation time}\label{sec:rot_cost}

Offline training for this flow would require the rotating DNS to be completed before the closure could be trained. Table~\ref{tab:rot_cost} compares the wall-clock time each run takes to the end of training with the time a periodic DNS takes to reach the same accuracy. Started from the perturbed $k$--$\omega$ solution, as it would be for a new flow, the periodic DNS must first develop its turbulence, and every embedded region shorter than the full length reaches the accuracy of table~\ref{tab:rot_jstar} sooner at both rotation numbers. Started from the developed non-rotating field, the periodic DNS is faster except for the half- and third-length regions at $Ro_b=0.5$. At $Ro_b=0.5$ the developed non-rotating field shortens the DNS $t^*$ little for the three longest regions, because the suction side must still relaminarise and because most of the time to their tight accuracies is spent averaging rather than in the transient. Against this start the $L_{x,0}/6$ and $L_{x,0}/8$ regions, although cheaper per step, need more updates, because their statistics are sampled where the flow is still adjusting from the rescaled inflow (section~\ref{sec:rot_length}). The inflow becomes rotation-aware only through feedback from the closure, and in these regions this takes more updates. The cheaper alternative, the baseline with a rotation correction, is less accurate than the full-, half- and third-length runs (table~\ref{tab:rot_jstar}). The advantage is expected to grow in domains spanning many integral scales, of which the embedded region covers only one or a few (section~\ref{sec:convergence_cost}).

\begin{table}
\centering
\begin{tabular}{llccccc}
\toprule
& & \multicolumn{5}{c}{$L_x/L_{x,0}=$} \\
$Ro_b$ & & $1$ & $1/2$ & $1/3$ & $1/6$ & $1/8$ \\
\midrule
0.15 & Error $J^*$ at end of training (table~\ref{tab:rot_jstar})      & 0.44 & 0.34 & 0.32 & 0.50 & 0.66 \\
     & oRANS & 0.25 & 0.12 & 0.051 & 0.046 & 0.042 \\
     & DNS from perturbed RANS solution                & 0.18 & 0.21 & 0.22 & 0.16 & 0.14 \\
     & DNS from developed non-rotating field           & 0.022 & 0.031 & 0.034 & 0.018 & 0.012 \\
\midrule
0.5  & Error $J^*$ at end of training (table~\ref{tab:rot_jstar})      & 0.19 & 0.16 & 0.13 & 0.50 & 0.72 \\
     & oRANS & 0.44 & 0.12 & 0.056 & 0.044 & 0.040 \\
     & DNS from perturbed RANS solution                & 0.25 & 0.28 & 0.32 & 0.13 & 0.091 \\
     & DNS from developed non-rotating field           & 0.21 & 0.26 & 0.31 & 0.023 & 0.0094 \\
\bottomrule
\end{tabular}
\caption{Wall-clock time $t^*$ of each oRANS run to the end of training and of the periodic DNS to reach the same accuracy, in units of the full reference DNS run ($t^*=1$, transient and averaging), as in figure~\ref{fig:convergence}. The first row of each block gives the accuracy of the oRANS run in that column, which is the target for the two DNS rows. The DNS is started either from the default RANS solution with random spanwise perturbations, as it would be for a new flow, or from the developed non-rotating field. The $L_x/L_{x,0}=1$ column is oRANS with a full-length embedded region, not the periodic DNS.}
\label{tab:rot_cost}
\end{table}

\subsection{Turbulence representation}\label{sec:rot_representation}

\begin{figure}
\centering
\includegraphics[width=\textwidth]{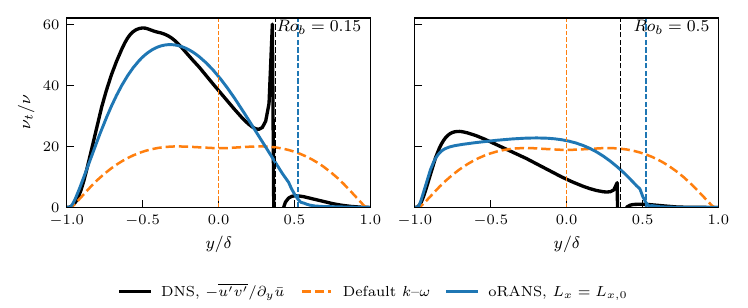}
\caption{Eddy viscosity for $Ro_b=0.15$ (left) and $0.5$ (right): DNS, evaluated under the Boussinesq assumption as $-\overline{u'v'}/\partial_y\bar u$; default $k$--$\omega$ and oRANS with $L_x=L_{x,0}$ at the end of training, $\alpha k/\omega$. Dashed lines mark the velocity maximum of each solution (DNS, black; default $k$--$\omega$, at the centreline; oRANS, blue), where the eddy viscosity evaluated from the DNS diverges.}
\label{fig:rot_nut}
\end{figure}

The closure learns the asymmetric eddy viscosity that rotation produces. Figure~\ref{fig:rot_nut} compares the eddy viscosity of the closures, $\nu_t=\alpha k/\omega$, with that of the DNS evaluated under the Boussinesq assumption, $\nu_t=-\overline{u'v'}/\partial_y\bar u$. In the non-rotating channel the eddy viscosity is symmetric. Under rotation it falls to the order of $\nu$ on the suction side, which carries almost no turbulent momentum flux, and rises above its non-rotating value on the pressure side. The baseline keeps the symmetric distribution, whereas the learned closure reduces $\nu_t$ to the DNS level on the suction side, which allows the velocity maximum to move, and recovers the pressure-side increase, more closely at $Ro_b=0.5$.

The production of turbulent kinetic energy, $P=-\overline{u'v'}\,\partial_y\bar u$, evaluated for the closures with the Boussinesq stress $-\overline{u'v'}=\nu_t\partial_y\bar u$, shows why this eddy viscosity reproduces the flow (table~\ref{tab:rot_budget}). In the rotating channel the mean shear in the core is set by the rotation, $\partial_y\bar u\simeq2\Omega$, so the core produces turbulence where the non-rotating channel produces almost none. At $Ro_b=0.5$ the baseline produces far too little there and the learned closure matches the DNS, which sustains the linear core. On the suction side the DNS production collapses and the learned closure follows it, while the baseline keeps its symmetric value. At $Ro_b=0.15$ the learned closure produces too little on the suction side, where the embedded region does not regenerate $k$ (section~\ref{sec:rot_length}).

\begin{table}
\centering
{\small\setlength{\tabcolsep}{4pt}
\begin{tabular}{llcccccc}
\toprule
& & \multicolumn{2}{c}{DNS} & \multicolumn{2}{c}{Default $k$--$\omega$} & \multicolumn{2}{c}{oRANS, $L_x=L_{x,0}$} \\
\cmidrule(lr){3-4}\cmidrule(lr){5-6}\cmidrule(lr){7-8}
$Ro_b$ & region & $P^+$ & $a_1$ & $P^+$ & $a_1$ & $P^+$ & $a_1$ \\
\midrule
0.15 & pressure side & 38  & 0.38 & 44  & 0.30 & 51  & 0.38 \\
     & core          & 4.2 & 0.27 & 1.6 & 0.16 & 6.2 & 0.35 \\
     & suction side  & 14  & 0.09 & 44  & 0.30 & 4.7 & 0.21 \\
\midrule
0.5  & pressure side & 49  & 0.31 & 41  & 0.30 & 59  & 0.34 \\
     & core          & 12  & 0.15 & 1.5 & 0.16 & 12  & 0.32 \\
     & suction side  & 3.1 & 0.04 & 41  & 0.30 & 1.2 & 0.16 \\
\bottomrule
\end{tabular}}
\caption{Production $P^+=P\,\nu/u_\tau^4$ and structure parameter $a_1$ (equation~\ref{eq:rot_a1}) of the DNS, the default $k$--$\omega$ model and the oRANS closure at the end of training, averaged over the pressure side ($-0.9<y/\delta<-0.3$), the core ($|y/\delta|<0.3$) and the suction side ($0.3<y/\delta<0.9$). $P^+$ is in units of $10^{-3}$.}
\label{tab:rot_budget}
\end{table}

The learned closure also responds to rotation in the classical way, seen in the structure parameter
\begin{equation}\label{eq:rot_a1}
a_1 = \frac{|\overline{u'v'}|}{k} = \frac{\nu_t\,|\partial_y\bar u|}{k},
\end{equation}
the efficiency with which turbulent energy is converted into shear stress, which takes the equilibrium value $\sqrt{C_\mu}\simeq0.3$ in a non-rotating logarithmic layer. As in the DNS, the learned $a_1$ rises above this value on the destabilised pressure side and falls below it on the stabilised suction side (table~\ref{tab:rot_budget}), where it remains too high for the same reason as the production deficit. Where the baseline is close to the DNS in table~\ref{tab:rot_budget}, in the pressure-side production and in the core value of $a_1$, its errors compensate, since its shear stress is too low and is offset by too high a mean shear and too low a $k$ respectively.

\begin{figure}
\centering
\includegraphics[width=\textwidth]{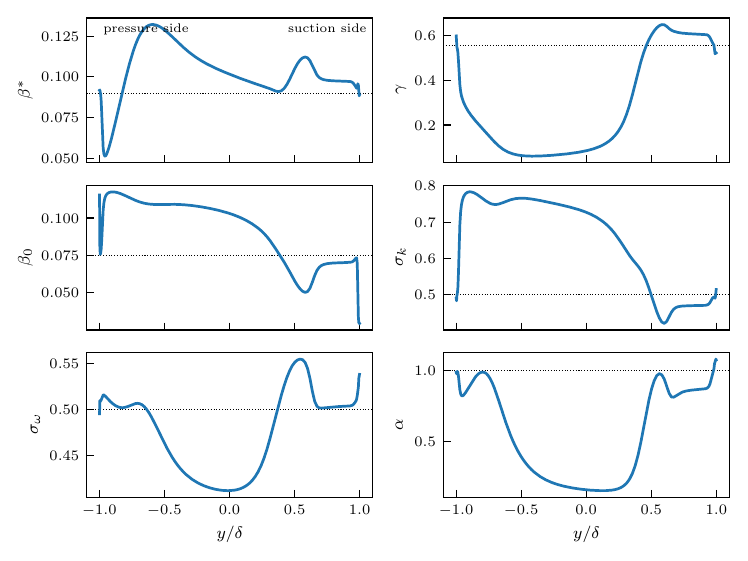}
\caption{Closure coefficients of the $Ro_b=0.5$, $L_x/L_{x,0}=1/2$ closure at the end of training ($J^*=0.16$). Pressure wall at $y/\delta=-1$. Dotted lines mark the standard $k$--$\omega$ values. Compare figure~\ref{fig:ransparameters} for the non-rotating channel.}
\label{fig:rot_coeffs}
\end{figure}

Finally, the coefficients show how the closure achieves this (figure~\ref{fig:rot_coeffs}, for the half-length embedded region at $Ro_b=0.5$). On the pressure side it keeps the near-wall structure of the channel closure of figure~\ref{fig:ransparameters}, with the minimum of $\beta^*$ near the wall and the log-law constraint satisfied. On the suction side, which is close to relaminarisation, neither appears. In the core the closure lowers $\alpha$, reducing the eddy viscosity where the mean shear is set by the rotation. The closure thus distinguishes the physics of the two sides of the channel from invariants that do not contain the rotation rate and from data generated inside the simulation.

\section{Conclusion}

We have introduced \textit{oRANS}, an online optimisation framework that couples a RANS solver with an embedded high-resolution sub-domain. By training directly on \emph{in situ} data from the target flow, oRANS circumvents the data-sparsity and over-fitting issues that limit offline machine-learning closures and is, in principle, applicable to any nonlinear PDE with unresolved terms. The method is validated on the stochastically forced Burgers equation, and in turbulent channel flow oRANS generally outperforms offline ML training, particularly in far out-of-sample test cases. In spanwise-rotating channel flow, where the Coriolis force vanishes from the RANS equations and a rotation-blind closure can only return the non-rotating solution, oRANS recovers the rotation-induced asymmetry from non-rotating inflow turbulence and a rotation-blind closure, whereas a closure trained offline on the non-rotating DNS recovers none of it. The learned closure is interpretable in both flows, recovering in the channel the near-wall damping of low-Reynolds-number models and outer values consistent with the logarithmic law, and learning in the rotating channel an eddy viscosity reduced to the order of the kinematic viscosity on the suction side, the production that rotation sustains in the core, and a structure parameter that responds as in the DNS. It maintains accuracy even for modest embedded domains where full periodic DNS may spuriously laminarise.

Beyond demonstrating proof of concept, the work contributes several methodological advances. We derived and implemented the discrete adjoint of a DL-augmented $k$--$\omega$ model for PDE-constrained optimisation, enabling efficient gradient computation within the RANS framework. A semi-implicit block-coupled RANS solver was developed with a stencil-wise reverse-mode implementation that preserves sparsity, yielding fast and stable time stepping for ML-augmented closures. Together, these advances establish a general framework for coupling low- and high-fidelity solvers that supports online training.

At the same time, we highlighted important limitations. The chief limitation is boundary-condition contamination. When the embedded domain is too short, spurious boundary effects contaminate the interior statistics, leading to self-reinforcing errors and, in extreme cases, algorithmic divergence. Exclusion of the lowest wavenumber modes for short domains also reduces accuracy, since the length of the embedded domain limits its ability to represent the largest turbulent motions. These challenges are particularly pronounced in turbulent channel flow, where simulations are often configured so that the largest turbulent structures are comparable to the domain size. In many applied flows, by contrast, the characteristic turbulent scales are much smaller than the overall domain size.

The results also show what the embedded region must supply. The RANS solution provides the mean velocity and the energy of the fluctuations at the inlet, the region regenerates the small and moderate scales, and a closure trained on it is as accurate as one trained with the largest scales of the flow present, until the region is too short to regenerate the moderate scales or its boundaries contaminate the statistics. The rotating-channel results set a condition for the corrective capability of the method. The embedded region must be long enough for the flow to relax from the model-conditioned inflow to the true dynamics before the statistics are sampled, since the training data are conditioned on the model being trained and the closure cannot learn what the region does not regenerate. The rotating channel shows that this condition can be met when the RANS solution supplying the inflow is qualitatively wrong. An inflow operator $\mathcal{T}$ that supplied the structure of the target turbulence, rather than rescaled amplitudes alone, would shorten this relaxation and is the natural next step.

When one (or several) embedded subdomains span the dominant physics, as is typical in quasi-homogeneous flows over many integral length scales, oRANS is expected to recover the full-domain low-order statistics at a cost set by the embedded region. The same optimisation strategy extends in principle to multi-block layouts and complex geometries, although accuracy is expected to decrease as the flow becomes strongly heterogeneous. These results support online optimisation with embedded data generation as a route to data-adaptive closures for flows in which the embedded region is representative of the turbulence elsewhere, and the formulation applies to other nonlinear PDEs with unclosed terms.

\section{Acknowledgments}
This work is supported by the U.K.\ Engineering and Physical Sciences Research Council grant EP/X031640/1 and the  U.S.\ National Science Foundation under Award CBET-22-15472. Daniel Dehtyriov's fellowship is supported by the Schmidt AI in Science Fellowship at the University of Oxford. This research used resources of the Oak Ridge Leadership Computing Facility at the Oak Ridge National Laboratory, which is supported by the Office of Science of the U.S.\ Department of Energy under Contract No.\ DE-AC05-00OR22725.

\appendix
\section{DNS Validation}\label{app:validation}

To validate the DNS solver, we reproduce the canonical turbulent channel flow dataset of \citet{Kim1987, Moser1999} at $Re_\tau \approx 180$. This case is a standard benchmark in the turbulence literature and provides both integral statistics and detailed turbulence profiles against which new solvers can be checked. Our simulation achieves a friction Reynolds number of $Re_\tau = 175$, close to the reference value. The simulation is driven at a fixed bulk velocity, so that $Re_b$ is prescribed and $Re_\tau$ is an outcome.

\begin{figure}
    \centering
    \includegraphics[width=0.8\textwidth]{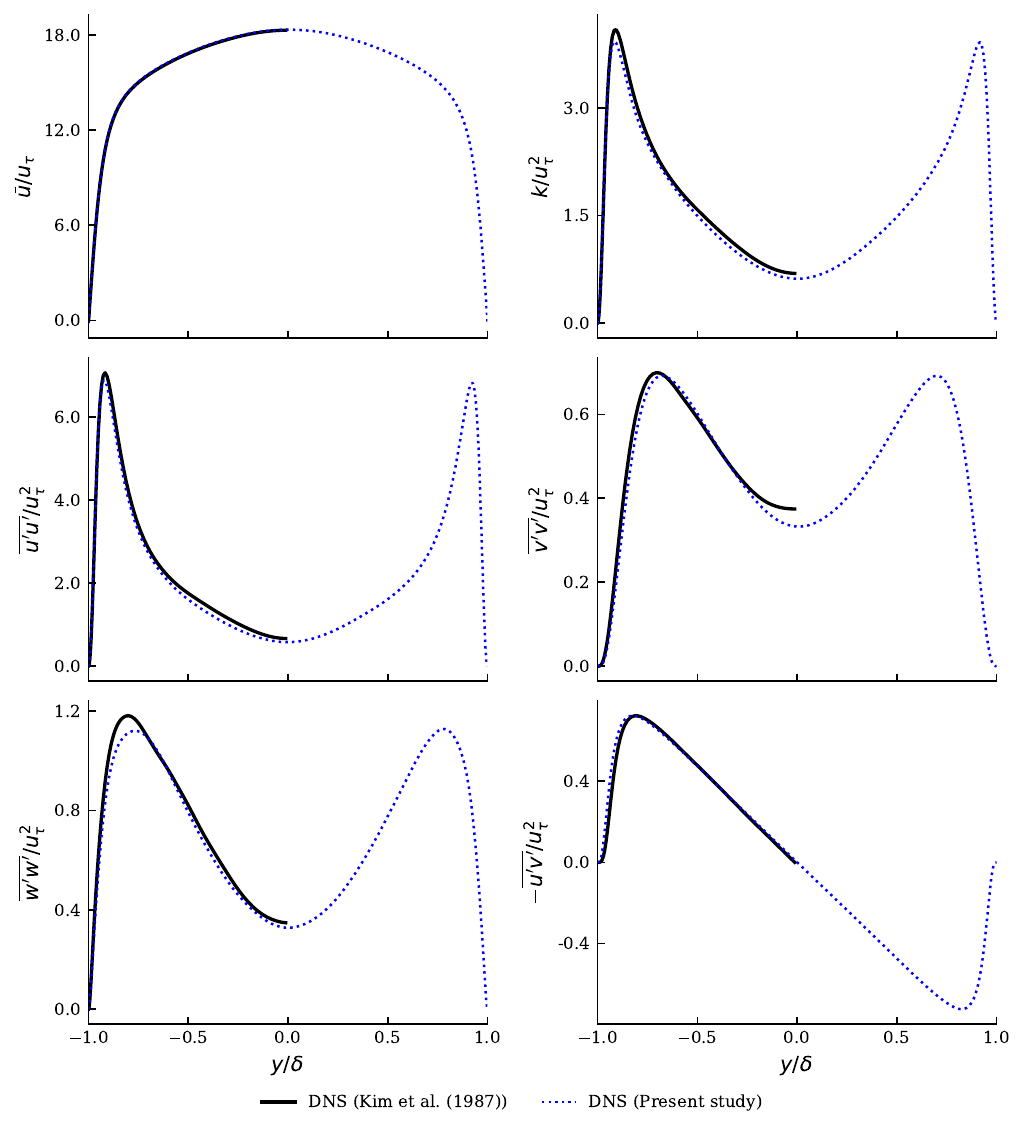}
    \caption{Validation of DNS implementation against the canonical turbulent channel flow dataset of \cite{Kim1987} for a target friction Reynolds number $Re_\tau = 180$. Profiles of mean velocity $\bar{u}/u_\tau$, turbulent kinetic energy $k/u_\tau^2$, and Reynolds stress components $\overline{u'u'}/u_\tau^2$, $\overline{v'v'}/u_\tau^2$, $\overline{w'w'}/u_\tau^2$, and $-\overline{u'v'}/u_\tau^2$ are shown. Present DNS results at $Re_\tau = 175$ (blue, dotted) are in close agreement with \cite{Kim1987} (black), confirming the accuracy of the simulation setup.}
    \label{fig:DNSvalidation}
\end{figure}

\begin{figure}
    \centering
    \includegraphics[width=\textwidth]{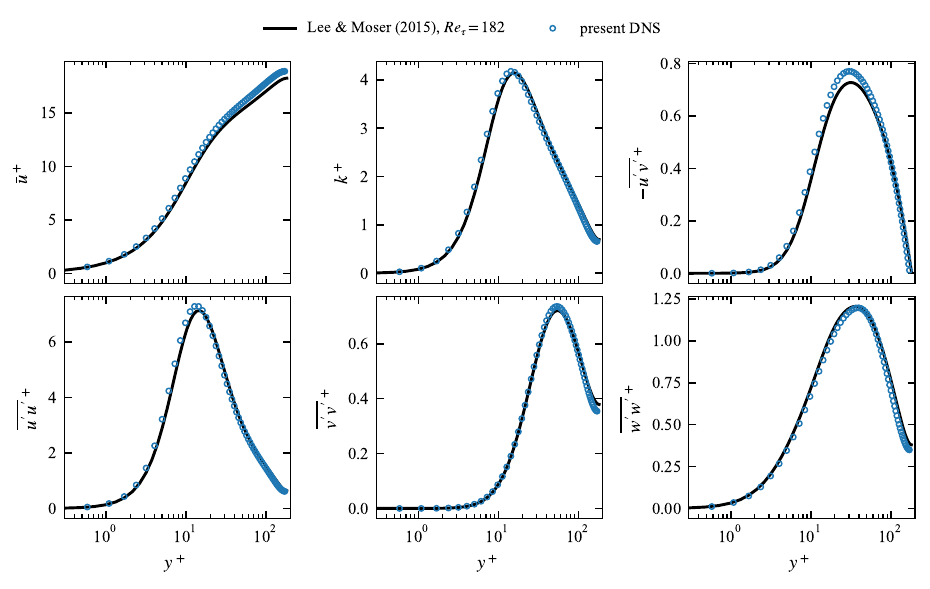}
    \caption{Validation of the DNS at $Re_\tau\simeq180$ against \citet{LeeMoser2015}: mean velocity, turbulent kinetic energy and the Reynolds stresses in wall units against $y^+$. Symbols, present DNS (both halves of the channel averaged); lines, \citet{LeeMoser2015} at $Re_\tau=182$.}
    \label{fig:DNSvalidationLM}
\end{figure}

Figure~\ref{fig:DNSvalidation} compares mean velocity, turbulent kinetic energy, and Reynolds stress components between our DNS and the KMM dataset, and figure~\ref{fig:DNSvalidationLM} makes the same comparison, in wall units, against the more recent and better-resolved simulation of \citet{LeeMoser2015} at $Re_\tau=182$. Overall agreement is excellent across the channel, with small differences attributable to the slightly lower $Re_\tau$ of our simulation. The near-wall peak in the streamwise stress and the location of the Reynolds shear-stress maximum are well captured. Minor discrepancies appear for the wall-normal stress $\overline{v'v'}$ at the channel centre which is slightly underpredicted, and the peak spanwise stress $\overline{w'w'}$ which is slightly overpredicted as compared to \citet{Kim1987}.

\begin{table}
\centering
\begin{tabular}{c|ccc}
\hline
Quantity & DNS & \cite{Kim1987} & \cite{LeeMoser2015} \\
\hline
$Re_c$        & 3,265 & 3,300 & 3,326 \\
$Re_b$        & 5,600 & 5,600 & 5,714 \\
$Re_\tau$     & 175   & 180   & 182 \\
$u_b/u_\tau$  & 16.15 & 15.63 & 15.69 \\
$C_f$         & $7.66 \times 10^{-3}$ & $8.18 \times 10^{-3}$ & $8.12 \times 10^{-3}$ \\
$u_c/u_b$     & 1.165 & 1.16 & 1.164 \\
$u_c/u_\tau$  & 18.81 & 18.20 & 18.26 \\
\hline
\end{tabular}
\caption{Comparison of channel flow statistics between the present DNS ($Re_\tau=175$) and the reference data of \cite{Kim1987} ($Re_\tau=180$) and \cite{LeeMoser2015} ($Re_\tau=182$). Bulk Reynolds number $Re_b$, friction Reynolds number $Re_\tau$, mean velocity ratio $u_b/u_\tau$, friction coefficient $C_f$, and centreline velocity $u_c$ ratios are reported. The close agreement across all quantities confirms that the present DNS accurately reproduces canonical turbulent channel flow at this Reynolds number.}
\label{tab:justin_kmm}
\end{table}

Against \citet{LeeMoser2015} the agreement is within a few per cent in every quantity, the peak of $\overline{u'u'}^+$ and the shear stress included, with the small differences in the outer region consistent with the difference in $Re_\tau$. Table~\ref{tab:justin_kmm} reports integral flow statistics. Bulk Reynolds number, velocity ratios, and the skin-friction coefficient all lie close to the reference values, confirming that the present solver reproduces canonical channel flow at this Reynolds number with high fidelity.

Together, these results establish that the present DNS implementation is consistent with benchmark data and provides a reliable high-fidelity reference for the oRANS framework.

\section{Rotating channel DNS validation}\label{app:rot_validation}

The rotating-channel DNS of section~\ref{sec:rotating} are validated
against the direct simulations of \citet{Kristoffersen1993} at the same
$Re_\tau=194$, which are the reference data for this flow. The domain,
resolution and numerical method are those of the non-rotating validation
above, with the Coriolis acceleration added to the momentum equations. The
rotation numbers are those of \citet{Kristoffersen1993}, $Ro_b=0.15$ and
$0.5$. Statistics are averaged over 800 samples
spanning $9.7\,\delta/u_\tau$, about twelve flow-through times, after a
transient of similar length, and the averaging window was verified to be
converged by comparing the statistics over its two halves. The sampling
periods of the reference are $9.1$, $5.6$ and $2.8\,\delta/u_\tau$ at
$Ro_b=0$, $0.15$ and $0.5$, which its authors note may leave some of their
statistics not fully converged.

\begin{figure}
    \centering
    \includegraphics[width=\textwidth]{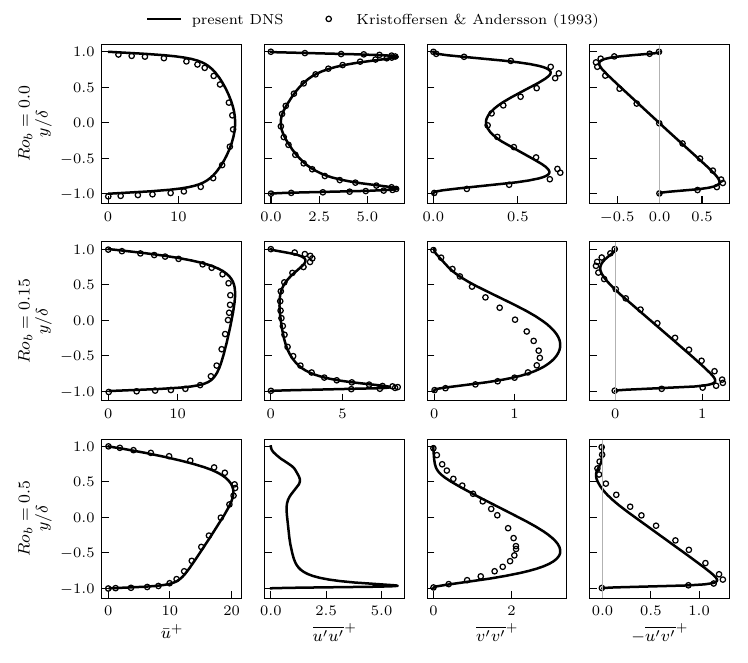}
    \caption{Validation of the rotating-channel DNS against
    \citet{Kristoffersen1993} at $Re_\tau=194$ for $Ro_b=0$ (top), $0.15$
    (centre) and $0.5$ (bottom): mean velocity, streamwise and wall-normal
    intensities and Reynolds shear stress in wall units based on the
    wall-averaged friction velocity. The pressure side is at
    $y/\delta=-1$. The reference streamwise intensity at $Ro_b=0.5$ could
    not be read reliably from the published figure and is omitted.}
    \label{fig:rot_validation}
\end{figure}

\begin{table}
\centering
\begin{tabular}{c|cc|cc|cc|cc}
\hline
$Ro_b$ & \multicolumn{2}{c|}{$u_b/u_\tau$} & \multicolumn{2}{c|}{$u_{\tau,p}/u_\tau$} & \multicolumn{2}{c|}{$u_{\tau,s}/u_\tau$} & \multicolumn{2}{c}{$\tau_{w,p}/\tau_{w,s}$} \\
 & present & K\&A & present & K\&A & present & K\&A & present & K\&A \\
\hline
0    & 15.55 & 15.06 & 0.951 & 1.002 & 0.954 & 1.004 & 0.99 & 1.00 \\
0.15 & 15.75 & 15.41 & 1.144 & 1.213 & 0.690 & 0.754 & 2.75 & 2.59 \\
0.5  & 15.30 & 15.25 & 1.164 & 1.207 & 0.594 & 0.679 & 3.84 & 3.16 \\
\hline
\end{tabular}
\caption{Integral quantities of the rotating-channel DNS at $Re_\tau=194$
and the corresponding values of \citet{Kristoffersen1993} (K\&A, their
table~3).}
\label{tab:rot_validation}
\end{table}

Figure~\ref{fig:rot_validation} compares the mean velocity, the streamwise
and wall-normal intensities and the Reynolds shear stress with the
reference data, and table~\ref{tab:rot_validation} the integral
quantities. The mean velocity, including the displacement of its maximum
towards the suction side and the linear core at $Ro_b=0.5$, the growth of
the pressure-side intensities and the collapse of the suction-side ones,
and the sign and magnitude of the shear stress across the channel are all
reproduced. The bulk velocity and the ratio of the wall shear stresses also
agree with the reference, least closely at $Ro_b=0.5$, the case the
reference authors sampled most briefly. The rotating DNS are therefore consistent with the benchmark and
provide a reliable reference for the rotating-channel results.

\bibliographystyle{elsarticle-harv}

\bibliography{references.bib}

\end{document}